\documentclass[
  journal=pasa,
  manuscript=research-article,
  year=2024,
  volume=XX,
]{cup-journal}

\usepackage{microtype,siunitx,booktabs}
\usepackage{makecell}
\usepackage{hyperref} 
\hypersetup{colorlinks,citecolor=blue,linkcolor=blue,urlcolor=blue}
\usepackage[nolist]{acronym}
\usepackage[version=3]{mhchem}
\usepackage{mathtools}
\usepackage{adjustbox}
\usepackage[normalem]{ulem}
\usepackage{xcolor}
\usepackage{comment}

\newcommand{\ergcm}[1]{erg\,cm$^{-2}$\,s$^{-1}$}
\def\HI{\hbox{H{\sc i}}}
\def\HII{\hbox{H{\sc ii}}}

\newcommand{\SII}{[S\,{\sc ii}]}
\newcommand{\OIII}{[O\,{\sc iii}]}

\newcommand{\Halpha}{H${\alpha}$}

\newcommand{\D}{$^\circ$}

\newcommand{\kms}{km\,s$^{-1}$}
\newcommand{\g}{G278.94+1.35}
\newcommand{\fs}{\mbox{\ensuremath{.\mkern-4mu^s}}}
\newcommand{\farcm}{\mbox{\ensuremath{.\mkern-4mu^\prime}}}
\newcommand{\farcs}{\mbox{\ensuremath{.\!\!^{\prime\prime}}}}
\newcommand{\fdg}{\mbox{\ensuremath{.\!\!^\circ}}}

\newcommand{\Ha}{H$\alpha$}

\def\HII{\hbox{H{\sc ii}}}
\def\HI{\hbox{H{\sc i}}}

\def\arcmin{\hbox{$^\prime$}}
\def\arcsec{\hbox{$^{\prime\prime}$}}
\def\kms{km\,s$^{-1}$}

\newcommand{\mjybm}{\,mJy\,beam$^{-1}$}
\newcommand{\ujybm}{\,$\mu$Jy\,beam$^{-1}$}

\newcommand{\um}{\,$\mu$m}

\begin{acronym}[AWGN]
\acro{2MASS}{Two Micron All Sky Survey}
\acro{2MASX}{Two Micron All Sky Survey Extended Source Catalogue}
\acro{30Dor}[30 Dor]{30~Doradus}
\acro{AGN}{active galactic nuclei}
\acro{ANR}{Agence Nationale de la Recherche}
\acro{ATCA}{Australia Telescope Compact Array}
\acro{ATESP}{Australia Telescope ESO Slice Project}
\acro{ATNF}{Australia Telescope National Facility}
\acro{ATOA}{Australia Telescope Online Archive}
\acro{AT20G}{Australia Telescope 20 GHz Survey}
\acro{ASKAP}{Australian Square Kilometre Array Pathfinder}
\acro{ARC}{Australian Research Council}
\acro{BETA}{Boolardy Engineering Test Array}
\acro{BL}[BL Lac]{BL Lacertae Objects}
\acro{CASS}{CSIRO Astronomy and Space Science}
\acro{CASDA}{CSIRO \ac{ASKAP} Science Data Archive}
\acro{CABB}{Compact Array Broadband Back-end}
\acro{CHII}[{\sc CHii}]{Compact \textsc{Hii}}
\acro{CMB}{Cosmic Microwave Background}
\acro{CSIRO}{Australian Commonwealth Scientific and Industrial Research Organisation}
\acro{CSS}{Compact Steep Spectrum}
\acro{DS9}[\textsc{DS9}]{\textsc{SAOImage DS9}}
\acro{DSA}{diffusive shock acceleration}
\acro{DSS}{Digital Sky Survey}
\acro{EM}{Electromagnetic}
\acro{EMU}{Evolutionary Map of the Universe}
\acro{eROSITA}{extended R\"Ontgen Survey with an Imaging Telescope Array}
\acro{ev}[eV]{electronvolt\acroextra{: 1 eV $\approx 1.6 \times 10^{-19}$ J}}
\acro{FITS}[\textsc{Fits}]{Flexible Image Transport System}
\acro{FRB}{Fast Radio Bursts}
\acro{FSRQ}{Flat Spectrum Radio Quasars}
\acro{FWHM}{Full Width at Half-Maximum}
\acro{GLEAM}{GaLactic Extragalactic All-sky MWA}
\acro{GPS}{Gigahertz Peak Spectrum}
\acro{HEMT}{High Electron Mobility Transistor}
\acro{HFP}[HFP]{High-Frequency Peaker}
\acro{HPBW}{Half Power Beam Width} 
\acro{HzRGs}{High Redshift Radio Galaxies}
\acro{pHFP}[pHFP]{Potential High Frequency Peaker}
\acro{HI}[H{\sc i}]{Neutral Atomic Hydrogen} 
\acro{HST}{\textit{Hubble Space Telescope}}
\acro{ICM}{intracluster medium}
\acro{IAU}{International Astronomical Union}
\acro{IFRSs}{Infrared Faint Radio Sources}
\acro{ISM}{interstellar medium}
\acro{IFRS}{Infrared Faint Radio Source}
\acro{IR}{Infrared}
\acro{JY}[Jy]{Jansky\acroextra{, 1 Jy = $10^{-26} \times \mathrm{W~ m}^{-2}~\mathrm{Hz}^{-1}$}} 
\acro{LLS}{Largest Linear Size}
\acro{LFAA}{Low-Frequency Aperture Array}
\acro{LMC}{Large Milky Cloud}
\acro{LOFAR}{Low-Frequency Array}
\acro{LSO}[LSOs]{Large Scale Objects}
\acro{MACHO}{Massive Astrophysical Compact Halo Objects}
\acro{MC}[MCs]{Magellanic Clouds}
\acro{mc}[MC]{Magellanic Cloud}    
\acro{MCELS}{Magellanic Cloud Emission Line Survey}
\acro{MW}{Milky Way}
\acro{MIRIAD}[\textsc{Miriad}]{Multichannel Image Reconstruction, Image Analysis and Display}
\acro{MIT}{Massachusetts Institute of Technology}
\acro{MOST}{Molonglo Observatory Synthesis Telescope}
\acro{MQS}{Magellanic Quasars Survey}
\acro{MRC}{Molonglo Reference Catalogue of Radio Sources}
\acro{MWA}{Murchison Widefield Array}
\acro{NRAO}{National Radio Astronomy Observatory}
\acro{NVSS}{NRAO VLA Sky Survey}
\acro{OPAL}{Online Proposal Applications \& Links}
\acrodefplural{ORC}{Odd Radio Circles}
\acro{ORC}{Odd Radio Circle}
\acro{OVV}{Optically Violent Variable Quasars}            
\acro{PAF}{Phased Array Feed}
\acro{pc}{parsec\acroextra{: 1 pc $\simeq 3.09 \times 10^{16}$ m}}
\acro{PMN}{Parkes-MIT-NRAO}
\acro{PNe}{Planetary Nebulae}
\acro{PWN}{Pulsar-wind Nebulae}
\acro{QSO}{Quasi-Stellar Object}
\acro{RA}{Right Ascension}
\acro{RFI}{Radio-Frequency Interference}
\acro{RMS}[rms]{root mean squared}
\acro{SARAO}{South African Radio Astronomy Observatory}
\acro{SDSS}{Sloan Digital Sky Survey}
\acro{SED}{spectral energy distribution}
\acro{SI}[$\alpha$]{Spectral Index\acroextra{, $S \propto \nu^\alpha$}}
\acro{SKA}{Square Kilometre Array}
\acro{SMB}{Super Massive Blackholes}
\acro{SMC}{Small Milky Cloud}
\acro{SN}{Supernova}
\acro{SUMSS}{Sydney University Molonglo Sky Survey}
\acro{TOPCAT}[\textsc{Topcat}]{Tool for OPerations on Catalogues And Tables}
\acro{USS}{Ultra Steep Spectrum}
\acro{YSOs}{young stellar objects}
\acro{WBAC}{Wide-Band Analogue Correlator}
\acro{WIFES}[WiFeS]{Wide-Field Spectrograph}
\acro{WISE}{Wide-Field Infrared Survey Explorer}
\acro{VLA}{Very Large Array} 
\acro{VLBI}{Very Long Baseline Interferometry} 
\acro{VLSR}[\textbf{$v_{lsr}$}]{Velocity in the Line of Sight}
\acro{SMASH}{Survey of the MAgellanic Stellar History}
\acro{SNR}{Supernova Remnant}
\acrodefplural{SNR}{Supernova Remnants}
\acro{SD}{single-degenerate}
\acro{SUMSS}{Sydney University Molonglo Sky Survey}
\acro{SMBH}{Super Massive Black Hole}
\end{acronym}

\title{Diprotodon on the sky. The Large Galactic \ac{SNR} \g}

\author{Miroslav D. Filipovi\'c}
\affiliation{Western Sydney University, Locked Bag 1797, Penrith South DC, NSW 2751, Australia}

\author{S. Lazarevi\'c}
\affiliation{Western Sydney University, Locked Bag 1797, Penrith South DC, NSW 2751, Australia}
\alsoaffiliation{CSIRO Space and Astronomy, Australia Telescope National Facility, PO Box 76, Epping, NSW 1710, Australia}
\alsoaffiliation{Astronomical Observatory, Volgina 7, 11060 Belgrade, Serbia}
\email[S. Lazarevi\'c]{S.Lazarevic@westernsydney.edu.au}

\author{M. Araya}
\affiliation{Escuela de F\'isica, Universidad de Costa Rica, San Jos\'e, 11501-2060, Costa Rica}

\author{N. Hurley-Walker}
\affiliation{International Centre for Radio Astronomy Research, Curtin University, Bentley, WA 6102, Australia}

\author{R. Kothes}
\affiliation{Dominion Radio Astrophysical Observatory, Herzberg Astronomy and Astrophysics, National Research Council Canada, PO Box 248, Penticton, BC V2A 6J9, Canada}

\author{H. Sano}
\affiliation{Faculty of Engineering, Gifu University, 1-1 Yanagido, Gifu 501-1193, Japan}
\alsoaffiliation{National Astronomical Observatory of Japan, Mitaka, Tokyo 181-8588, Japan}

\author{G. Rowell}
\affiliation{School of Physics, Chemistry, and Earth Sciences, The University of Adelaide, Adelaide 5005, Australia} 

\author{P. Martin}
\affiliation{IRAP, Universit\'e de Toulouse, CNRS, CNES, F-31028 Toulouse, France}

\author{Y. Fukui}
\affiliation{Department of Physics, Nagoya University, Furo-cho, Chikusa-ku, Nagoya 464-8601, Japan}

\author{R. Z. E. Alsaberi}
\affiliation{Western Sydney University, Locked Bag 1797, Penrith South DC, NSW 2751, Australia}

\author{B. Arbutina}
\affiliation{Department of Astronomy, Faculty of Mathematics, University of Belgrade, Studentski trg 16, 11000 Belgrade, Serbia}

\author{B. Ball}
\affiliation{Department of Physics, University of Alberta, 4-181 CCIS, Edmonton, Alberta T6G 2E1, Canada}

\author{C. Bordiu}
\affiliation{INAF\,$-$\,Osservatorio Astrofisico di Catania, Via S. Sofia 78, I-95123, Catania, Italy} 

\author{R. Brose}
\affiliation{School of Physical Sciences and Centre for Astrophysics \& Relativity, Dublin City University, D09 W6Y4 Glasnevin, Ireland}
\alsoaffiliation{Dublin Institute for Advanced Studies, Astronomy \& Astrophysics Section, DIAS Dunsink Observatory, Dublin, D15 XR2R, Ireland}

\author{F. Bufano}
\affiliation{INAF\,$-$\,Osservatorio Astrofisico di Catania, Via S. Sofia 78, I-95123, Catania, Italy} 

\author{C. Burger-Scheidlin}
\affiliation{Dublin Institute for Advanced Studies, Astronomy \& Astrophysics Section, DIAS Dunsink Observatory, Dublin, D15 XR2R, Ireland}
\alsoaffiliation{School of Physics, University College Dublin, Belfield, Dublin, D04 V1W8, Ireland}

\author{T. A. Collins}
\affiliation{Institut für Physik und Astronomie, Universität Potsdam, Karl-Liebknecht-Strasse 24/25, D 14476 Potsdam, Germany} 

\author{E. J. Crawford}
\affiliation{Western Sydney University, Locked Bag 1797, Penrith South DC, NSW 2751, Australia}

\author{S. Dai}
\affiliation{Western Sydney University, Locked Bag 1797, Penrith South DC, NSW 2751, Australia}

\author{S. W. Duchesne}
\affiliation{CSIRO Space and Astronomy, PO Box 1130, Bentley WA 6102, Australia}

\author{R. S. Fuller}
\affiliation{Western Sydney University, Locked Bag 1797, Penrith South DC, NSW 2751, Australia}

\author{A. M. Hopkins}
\affiliation{School of Mathematical and Physical Sciences, 12 Wally's Walk, Macquarie University, NSW 2109, Australia}

\author{A. Ingallinera}
\affiliation{INAF\,$-$\,Osservatorio Astrofisico di Catania, Via S. Sofia 78, I-95123, Catania, Italy}

\author{H. Inoue}
\affiliation{Faculty of Engineering, Gifu University, 1-1 Yanagido, Gifu 501-1193, Japan}

\author{T. H. Jarrett}
\affiliation{Department of Astronomy, University of Cape Town, Private Bag X3, Rondebosch 7701, South Africa}
\alsoaffiliation{Western Sydney University, Locked Bag 1797, Penrith South DC, NSW 2751, Australia}

\author{B. S. Koribalski}
\affiliation{CSIRO Space and Astronomy, Australia Telescope National Facility, PO Box 76, Epping, NSW 1710, Australia}
\alsoaffiliation{Western Sydney University, Locked Bag 1797, Penrith South DC, NSW 2751, Australia}

\author{D. Leahy}
\affiliation{Department of Physics and Astronomy, University of Calgary, Calgary, Alberta, T2N 1N4, Canada}

\author{K. J. Luken}
\affiliation{Western Sydney University, Locked Bag 1797, Penrith South DC, NSW 2751, Australia}
\alsoaffiliation{Data61, CSIRO, Epping, NSW, Australia}

\author{J. Mackey}
\affiliation{Dublin Institute for Advanced Studies, Astronomy \& Astrophysics Section, DIAS Dunsink Observatory, Dublin, D15 XR2R, Ireland}
\alsoaffiliation{School of Physics, University College Dublin, Belfield, Dublin, D04 V1W8, Ireland}

\author{P. J. Macgregor}
\affiliation{Western Sydney University, Locked Bag 1797, Penrith South DC, NSW 2751, Australia}
\alsoaffiliation{CSIRO Space and Astronomy, Australia Telescope National Facility, PO Box 76, Epping, NSW 1710, Australia}

\author{R. P. Norris}
\affiliation{Western Sydney University, Locked Bag 1797, Penrith South DC, NSW 2751, Australia}
\alsoaffiliation{CSIRO Space and Astronomy, Australia Telescope National Facility, PO Box 76, Epping, NSW 1710, Australia}

\author{J. L. Payne}
\affiliation{Western Sydney University, Locked Bag 1797, Penrith South DC, NSW 2751, Australia}

\author{S. Riggi}
\affiliation{INAF\,$-$\,Osservatorio Astrofisico di Catania, Via S. Sofia 78, I-95123, Catania, Italy} 

\author{C.~J.~Riseley}
\affiliation{Dipartimento di Fisica e Astronomia, Università degli Studi di Bologna, via P. Gobetti 93/2, 40129 Bologna, Italy}
\alsoaffiliation{INAF -- Istituto di Radioastronomia, via P. Gobetti 101, 40129 Bologna, Italy}

\author{M. Sasaki}
\affiliation{Dr Karl Remeis Observatory, Erlangen Centre for Astroparticle Physics, Friedrich-Alexander-Universit\"{a}t Erlangen-N\"{u}rnberg, Sternwartstra{\ss}e 7, 96049 Bamberg, Germany}

\author{Z. J. Smeaton}
\affiliation{Western Sydney University, Locked Bag 1797, Penrith South DC, NSW 2751, Australia}

\author{I. Sushch}
\alsoaffiliation{CIEMAT, Avda. Complutense 40, 28040 Madrid, Spain}
\affiliation{Gran Sasso Science Institute, Via F.Crispi 7, 67100 L’Aquila, Italy}
\alsoaffiliation{INFN-Laboratori Nazionali del Gran Sasso, Via G. Acitelli 22, Assergi (AQ), Italy}
\alsoaffiliation{Astronomical Observatory of Ivan Franko National University of Lviv, Kyryla i Methodia 8, UA-79005 Lviv, Ukraine}
\alsoaffiliation{Centre for Space Research, North-West University, 2520 Potchefstroom, South Africa} 

\author{M. Stupar}
\affiliation{Western Sydney University, Locked Bag 1797, Penrith South DC, NSW 2751, Australia}

\author{G. Umana}
\affiliation{INAF\,$-$\,Osservatorio Astrofisico di Catania, Via S. Sofia 78, I-95123, Catania, Italy} 

\author{D. Uro\v{s}evi\'c}
\affiliation{Department of Astronomy, Faculty of Mathematics, University of Belgrade, Studentski trg 16, 11000 Belgrade, Serbia}

\author{V. Velovi\'c}
\affiliation{Western Sydney University, Locked Bag 1797, Penrith South DC, NSW 2751, Australia}

\author{T. Vernstrom}
\affiliation{ICRAR, The University of Western Australia, 35 Stirling Hw, 6009 Crawley, Australia}

\author{B. Vukoti\'c}
\affiliation{Astronomical Observatory, Volgina 7, 11060 Belgrade, Serbia}

\author{J. West}
\affiliation{Dominion Radio Astrophysical Observatory, Herzberg Astronomy and Astrophysics, National Research Council Canada, PO Box 248, Penticton, BC V2A 6J9, Canada}

\doi{}

\received {dd Mmm YYYY}
\revised  {dd Mmm YYYY}
\accepted {dd Mmm YYYY}
\published{dd Mmm YYYY}


\begin{document}

\begin{abstract}
We present a re-discovery of \g\footnote{Also known as G279.0+1.1 \citep{Green}.} as possibly one of the largest known Galactic supernova remnants (\acp{SNR}) -- that we name Diprotodon. While previously established as a Galactic \ac{SNR}, Diprotodon is visible in our new \ac{EMU} and \ac{GLEAM} radio continuum images at an angular size of 3\fdg33$\times$3\fdg23, much larger than previously measured. At the previously suggested distance of 2.7\,kpc, this implies a diameter of 157$\times$152\,pc. This size would qualify Diprotodon as the largest known \ac{SNR} and pushes our estimates of \ac{SNR} sizes to the upper limits. We investigate the environment in which the \ac{SNR} is located and examine various scenarios that might explain such a large and relatively bright \ac{SNR} appearance. We find that Diprotodon is most likely at a much closer distance of $\sim$1\,kpc, implying its diameter is 58$\times$56\,pc and it is in the radiative evolutionary phase. We also present a new \textit{Fermi}-LAT data analysis that confirms the angular extent of the \ac{SNR} in gamma-rays. The origin of the high-energy emission remains somewhat puzzling, and the scenarios we explore reveal new puzzles, given this unexpected and unique observation of a seemingly evolved \ac{SNR} having a hard GeV spectrum with no breaks. We explore both leptonic and hadronic scenarios, as well as the possibility that the high-energy emission arises from the leftover particle population of a historic pulsar wind nebula.
\end{abstract}

\acresetall 

\section{Introduction}
\label{sec:introduction}

\acp{SNR} are essential ingredients in the evolution of every galaxy, as they are one of the main sources of \ac{ISM} enrichment. They significantly impact the structure and physical properties of the surrounding \ac{ISM} \citep{book2}. It is well understood that the census of the Galactic \ac{SNR} population is incomplete \citep{2013A&A...549A.107F,2021A&A...651A..86D,2023MNRAS.524.1396B}. Some 300+ such objects are confirmed \citep{Green,Ferrand2012} with the expectation that up to $\sim$2000 additional \acp{SNR} remain undiscovered in the Milky Way \citep{2022ApJ...940...63R}. As recently shown in \citet{2023MNRAS.524.1396B}, a significant number of the missing Galactic \acp{SNR} are expected to have a low-surface brightness or be located in complex regions where clear distinctions from other source types (e.g. \HII~regions) are challenging. It is also well known that new, bright, small-sized (compact) and presumably young \acp{SNR} are not likely to be found in abundance \citep{2021Univ....7..338R} and Smeaton et al. (2024, in press). 

At the same time, among the 300-strong Galactic \ac{SNR} population, there are a significant number of these objects whose measured position and extent are based on poorer resolution observations. This is due to the observations being conducted using the previous generation of instruments, such as the \ac{VLA}, Parkes/Murriyang, the \ac{ATCA}, Effelsberg, \ac{MOST}, and the synthesis telescope at the Dominion Radio Astrophysical Observatory (DRAO~ST) as part of the Canadian Galactic Plane Survey \citep[CGPS,][]{cgps}. Thus, the newer generation of radio telescopes, such as \ac{ASKAP} \citep{2007PASA...24..174J,Norris2011,2021PASA...38...46N,2021PASA...38....9H,2022arXiv220808245K}, MeerKAT \citep{2016mks..confE...1J}, \ac{MWA} \citep{2019PASA...36...50B}, and \ac{LOFAR} \citep{2013A&A...556A...2V}, are imperative in improving these previous measurements, as well as discovering new Galactic \acp{SNR}.

As shown in several recent studies, discoveries such as the intergalactic \ac{SNR} J0624--6948 \citep{2022MNRAS.512..265F}, the Galactic \acp{SNR} G288.8--6.3 \citep{2023AJ....166..149F,2024A&A...684A.150B}, G181.1--9.5 \citep{2017A&A...597A.116K}, Hoinga \citep{2021A&A...648A..30B}, G118.4+37.0 \citep[Calvera;][]{2022A&A...667A..71A}, J1818.0--1607 \citep{2023ApJ...943...20I}, G308.73+1.38 \citep[Raspberry;][]{2024RNAAS...8..107L}, G312.65+2.87 \citep[Unicycle;][]{2024RNAAS...8..158S} and G121.1--1.9 \citep{2023MNRAS.521.5536K} demonstrate the ability of these newer telescopes to discover new Galactic \acp{SNR}. 
These \acp{SNR} are mainly located well outside the Galactic Plane, where they can preserve their original circular \ac{SNR} shape for longer timeframes due to the presumably low-density environment while also displaying a lower surface-brightness as compared to typical \acp{SNR}. 

\g, here named Diprotodon (see~\ref{sec:app}), is a Galactic \ac{SNR} originally established by \citet{1988MNRAS.234..971W} using the South African 26-m Hartesbeesthoek radio telescope at 1.6\,GHz (restoring B.S.\footnote{Beam Size.}=30\arcmin; S=25.2$\pm$4\,Jy) and 2.3\,GHz (B.S.=20\arcmin; S=20.7$\pm$3\,Jy). \citet{1995MNRAS.277..319D} used the Parkes 64-m radio-telescope at 1.4\,GHz (B.S.=18\arcmin; S=28.3$\pm$3\,Jy) and 2.4\,GHz (B.S.=11\arcmin; S=20$\pm$2\,Jy) to study the morphology and polarisation of Diprotodon. Finally, \citet{1996A&AS..118..329W} detected some parts of Diprotodon in the MOST survey at 843\,MHz (B.S.=43\arcsec$\times$53\arcsec). All these radio studies confirmed Diprotodon as having a faint and incomplete shell.
Diprotodon has also been detected and confirmed optically with \Ha\ \citep{2009MNRAS.394.1791S,2011MNRAS.414.2282S}, as well as in gamma-rays in the energy range 0.5--500\,GeV, which revealed a source larger than the then known radio shell \citep{2020MNRAS.492.5980A}. Several studies investigated if any of the pulsars that are found in the vicinity of Diprotodon could be associated with the \ac{SNR} itself, but no convincing relationship between a pulsar and Diprotodon was established, including PSR\,J0940-5428 \citep{2024A&A...685A..23M}. Most recently, \citet{2024A&A...685A..23M} presented eROSITA's detection of Diprotodon's X-ray counterpart showing that the X-ray emission is soft, coming from a narrow range of energies between 0.3 and 1.5\,keV.

The original distance of 3\,kpc was established using Parkes 64-m OH observations \citep{1997AJ....114.2058G}. \citet{2019RAA....19...92S} refined this value to 2.7$\pm$0.3\,kpc using optical extinction. 
This was determined with the red clump star method, which could be somewhat unreliable as uncertainties are typically highly underestimated. With realistic uncertainties considered, this distance estimate should have a much larger error. \citet{2024A&A...685A..23M} also considered a distance of $\sim$400\,pc that could be associated with local/nearby pulsars.
With this caution in mind, and in the absence of any more reliable distance estimate, we initially adopt 2.7$\pm$0.3\,kpc as the assumed distance to Diprotodon. However, after in-depth analysis (Section~\ref{sec:results}), we argue that the most likely distance to this \ac{SNR} is at $\sim$1\,kpc. 

\section{Observations and Data Processing}
\label{sec:observations}

\subsection{Radio Observations}

\subsubsection{ASKAP data}
\label{askap}

\begin{figure*}[t!]
\centering
    \includegraphics[trim=0 10 0 0,width=\linewidth]{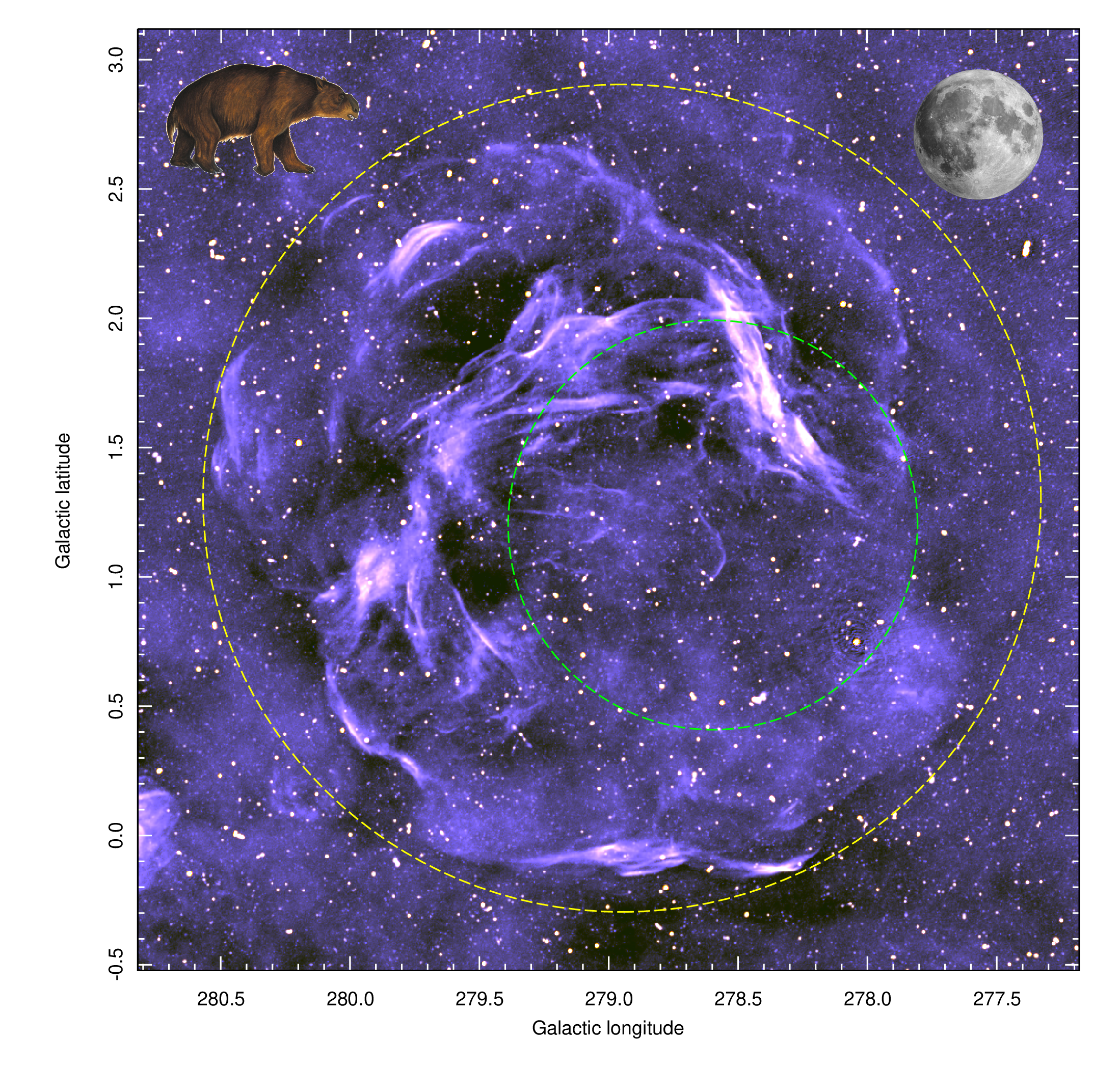}
    \caption{\ac{ASKAP} radio-continuum intensity image of Diprotodon at 943\,MHz. The green-dashed, 95\arcmin--diameter circle indicates the previously measured extent, while the yellow dash ellipse indicates the new boundaries of Diprotodon's radio emission (3\fdg33$\times$3\fdg23). In the top right corner, we show the scaled size of the Moon (0\fdg5), while in the top left corner, we show the animal Diprotodon. 
    }
    \label{fig:1}
\end{figure*}

Our serendipitous detection of Diprotodon was enabled by the \ac{ASKAP}-\ac{EMU} project (AS201), which observed this area of the radio sky in July 2023 with a complete set of 36 \ac{ASKAP} antennas at the central frequency of 943.4\,MHz and bandwidth of 288~MHz. All data are available through the \ac{CSIRO} \ac{ASKAP} Science Data Archive (CASDA\footnote{\url{https://research.csiro.au/casda}}). The observations containing this object are the tiles EMU\_0954$-$55 and 1005--51 corresponding to \ac{ASKAP} scheduling blocks SB51428 and SB54774. The data were processed using the ASKAPsoft pipelines (SB51428 was made with V.2.5.18 and SB54774 was with V.2.8.3.), which included multi-frequency synthesis imaging, multi-scale cleaning, self-calibration and convolution to a common beam size. \citep{askapsoft_2019ascl.soft12003G}. The resulting 943~MHz \ac{EMU} image has a \ac{RMS} sensitivity of $\sigma$=25\ujybm\ and a synthesised beam of 15\arcsec$\times$15\arcsec\ (Figure~\ref{fig:1}).

\subsubsection{GLEAM data}
 \label{sec:GLEAM}

\begin{figure*}[t!]
\centering
  \includegraphics[width=\linewidth,trim=5 40 0 35,clip]{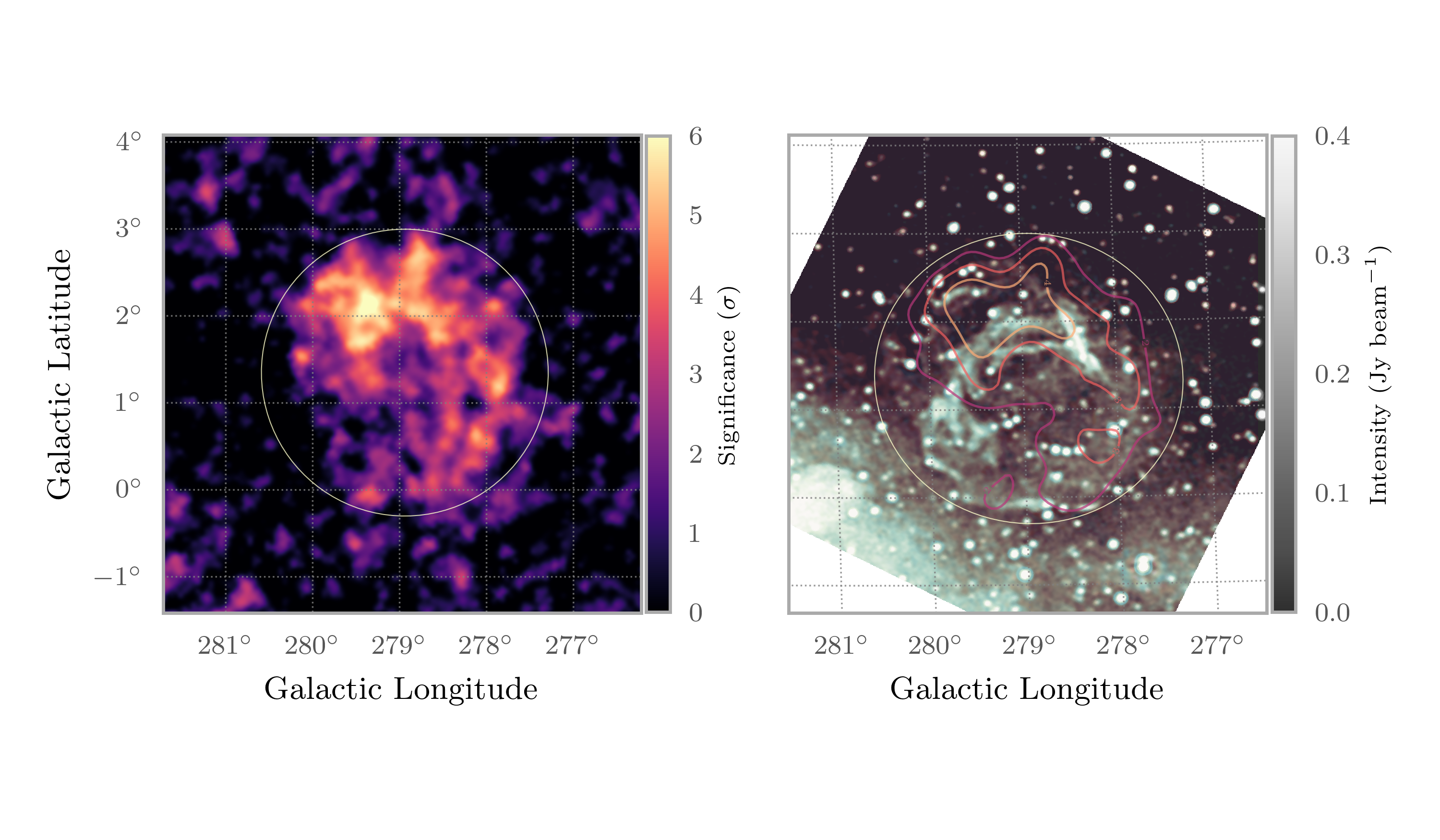} 
    \caption{   
    Left: Significance map (in units of $\sigma$) of the gamma-ray emission from Diprotodon obtained with \textit{Fermi}-LAT events for energies above 1\,GeV. 
    Right: \ac{GLEAM} RGB intensity image of Diprotodon where R is (red) at 88\,MHz, B (blue) at 118\,MHz and G (green) at 154\,MHz. The contours correspond to 2, 3, and 4~$\sigma$ significance levels shown in the \textit{Fermi}-LAT map on the left.
    The circle represents the measured diameter of Diprotodon of 3\fdg33$\times$3\fdg23.
    }
    \label{fig:gleam}
\end{figure*}

We examined several radio surveys to search for Diprotodon and to derive the flux density as a function of frequency. Only surveys with sensitivity to scales at least that of its diameter (3\fdg33) can be used so that it is not resolved out. At the lowest frequency, we used data taken by the \ac{MWA} \citep[][]{2013PASA...30....7T,2018PASA...35...33W} for the GaLactic and All-sky \ac{MWA} \citep[GLEAM;][]{2015PASA...32...25W} survey, at the edge of the 103--134\,MHz source-finding mosaics generated by \cite{2017MNRAS.464.1146H} (Figure~\ref{fig:gleam}). The resolution of this 118-MHz image is 3\farcm2$\times$3\farcm2 and the \ac{RMS} noise is 55\,\mjybm\ (measured from a patch with no sources and little Galactic emission).

To remove the contaminating point sources from the data, we followed established methods  \citep{2005A&A...436..187T,2021A&A...648A..30B,2022MNRAS.510.2920A,2023MNRAS.524.1396B}. 
We performed source-finding on the \ac{MWA} image, using \textsc{Aegean}\footnote{\url{https://github.com/PaulHancock/Aegean}} \citep{2012MNRAS.422.1812H,2018PASA...35...11H} and its companion tool, the Background and Noise Estimator (\textsc{BANE}). 
We subtracted the \ac{MWA} point sources from the image using a further ancillary tool \textsc{AeRes}, and from the resulting image, we measured Diprotodon's flux density (Section~\ref{sec:radio_SED}).

\subsubsection{HI4PI, SGPS and NANTEN data}
 \label{sec:other}
 
We also searched for possible detection of \HI\ and CO clouds associated with Diprotodon using the Effelsberg and Parkes HI4PI survey \citep{2016A&A...594A.116H} and the NANTEN $^{12}$CO($J$~=~1--0) data \citep{2004ASPC..317...59M}. The angular resolutions of \HI\ and CO data are $\sim$16\farcm2 and $\sim$2\farcm6, respectively. The typical noise fluctuation is $\sim$0.04~K at the velocity resolution of 1.3\,\kms\ for the \HI\ data, and $\sim$0.2~K at the velocity resolution of 1\,\kms\ for the CO data. Details of these results are presented in Section~\ref{sec:radio_HI}.

The southern part of Diprotodon -- seen in Galactic coordinates -- is covered by the Southern Galactic Plane Survey \citep[SGPS,][]{2005ApJS..158..178M}. The SGPS contains a \HI\ spectral line survey covering Galactic longitudes from {253\D} to {358\D} and latitudes from --1\fdg5 to +1\fdg5. The survey combines observations with the \ac{ATCA} and the Parkes telescopes, giving a spatial resolution of $\sim$2\arcmin.

\subsection{\textit{Fermi}-LAT Observations}
  \label{sec:Fermi}

The \textit{Fermi} Large Area Telescope (LAT) is a converter/tracker observatory detecting photons at high energies from $\sim$20~MeV to $>$1~TeV and surveying the entire sky since 2008 \citep{2009ApJ...697.1071A}. We used Pass~8 data collected from 04 August 2008 to 30 June 2023 in the energy range of 0.2\,GeV--500\,GeV. We used {\tt fermitools} version~2.2.0 through the {\tt fermipy} package version~1.1.6 to analyse the data. We selected front and back-converted events with good quality in the {\tt SOURCE} class using the options {\tt evclass=128}, {\tt evtype=3}, {\tt DATA\_QUAL$>$0}, and having zenith angles lower than $90^\circ$ to avoid contamination from gamma rays from Earth's limb. We used the corresponding response functions {\tt P8R3\_SOURCE\_V3} and binned the data with a spatial scale of 0\fdg05 per pixel and ten bins in energy for exposure calculation. 

The analysis of LAT data is based on the maximum likelihood technique \citep{1996ApJ...461..396M} by which the morphological and spectral parameters of the sources in the region of interest (RoI) are fit to account for the number of events in each spatial and energy bin. Given the relatively large point-spread function of the LAT, we collected events from a $20^\circ \times 20^\circ$ RoI centred at the coordinates RA(J2000)\,=\,10$^{\rm h}$00$^{\rm m}$00\fs0 and Dec(J2000)\,=\,--53\D00\arcmin00\farcs0 and included in the model all sources located within 25$^\circ$ of the RoI centre found in the latest incremental version of the \textit{Fermi} Large Area Telescope Fourth Source Catalog \citep[4FGL-DR4,][]{2020ApJS..247...33A,2023arXiv230712546B}, based on the first 14 years of data. We also included in the model the Galactic diffuse emission using the file {\tt gll\_iem\_v07.fits} and the isotropic and residual cosmic-ray background as described by {\tt iso\_P8R3\_SOURCE\_V3\_v1.txt}\footnote{\url{https://fermi.gsfc.nasa.gov/ssc/data/access/lat/BackgroundModels.html}}. Energy dispersion correction was not applied to the isotropic component as recommended by the LAT team\footnote{\url{https://fermi.gsfc.nasa.gov/ssc/data/analysis/documentation/Pass8\_edisp\_usage.html}}. In the 4FGL-DR4 catalogue, the emission associated with Diprotodon is described by a spatial template based on the analysis by \cite{2020MNRAS.492.5980A} and labelled 4FGL~J1000.0--5312e. The source is described by a ring with an outer diameter of 2\fdg88.

The detection significance of a source (in $\sigma$) having one additional free parameter in the model is obtained from the square root of the test statistic (TS), defined as $-2\log(\mathcal{L}_0/\mathcal{L})$, with $\mathcal{L}$ and $\mathcal{L}_0$ the maximum likelihood functions for a model including a source and for the model without this additional source, respectively. To maximize the likelihood in the initial step, we fit the spectral normalisation of the sources found within $10^\circ$ of the RoI centre as well as all the spectral parameters of the sources found within $5^\circ$ using events with energies above 0.2\,GeV. In the next step, we improved the background model by adding new point sources at the locations of TS maxima exceeding a value of 16 in the residual emission obtained with the standard 4FGL-DR4 model. This was particularly important at the lowest energies of the analysis because of the presence of considerable background gamma-ray excesses in the RoI. In the next fit, we optimised the values of the spectral parameters of the new sources located up to $12^\circ$ from the centre. For the rest of the analysis of the spectrum and morphology of Diprotodon, due to a large number of free parameters, we only kept free the spectral normalisations of the sources located within $8^\circ$ and the spectral indices and normalisations of the sources located within $5^\circ$ of the centre.

To visualize the gamma-ray emission from the \ac{SNR}, we calculated a significance map for events having energies above 1\,GeV. We removed the source 4FGL~J1000.0--5312e from the model and placed a putative point source at each pixel in the map to fit its spectral normalisation (in this case, the additional parameter). The differential spectral model used for the point source is a simple power-law with a fixed index of 2. The resulting significance map is seen in Figure~\ref{fig:gleam}, clearly showing the \ac{SNR}.

\subsection{WISE View of Diprotodon}
 \label{sec:IR}

\begin{figure*}[t!]
\centering
    \includegraphics[scale=0.95,clip]{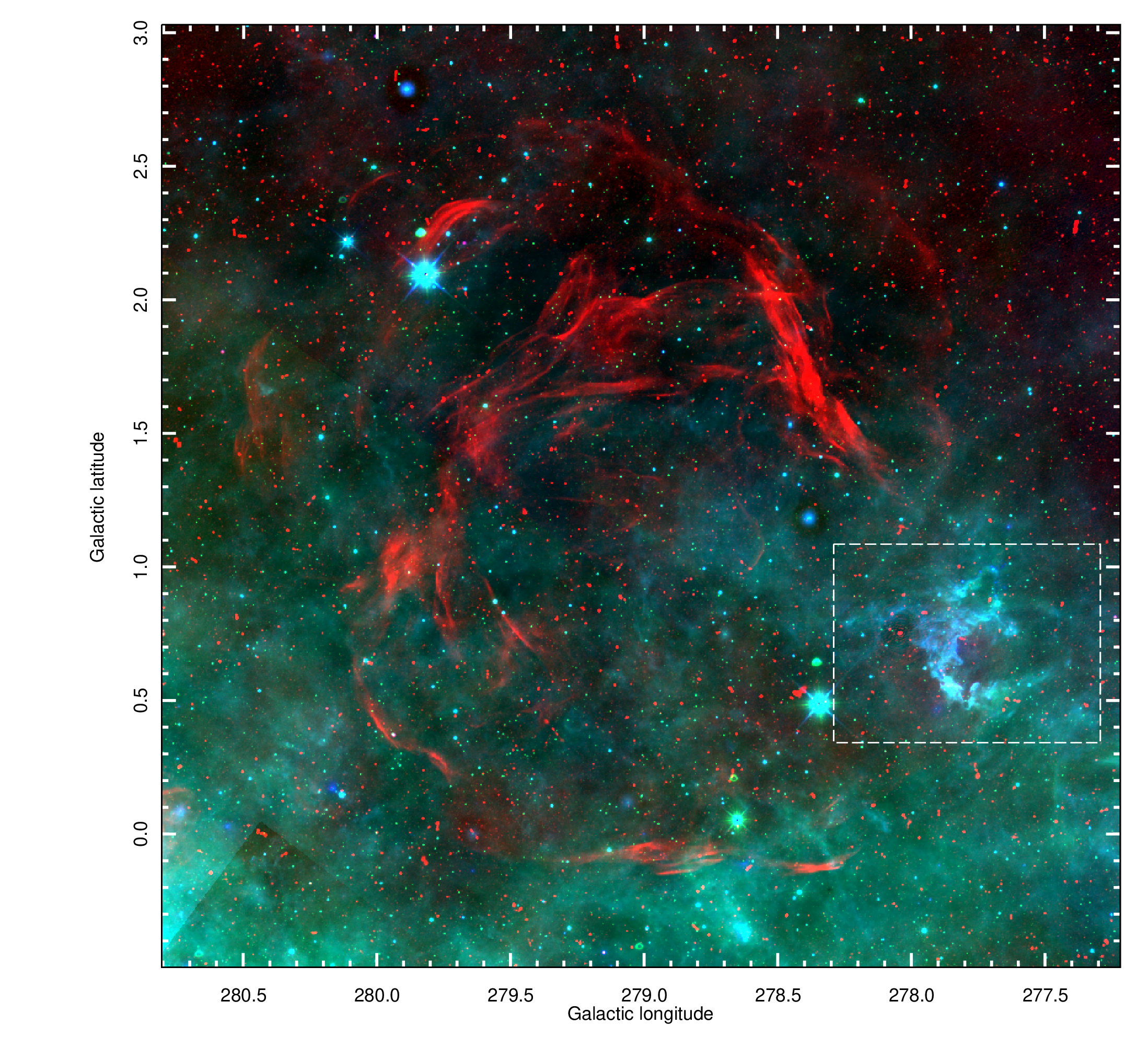}
    \caption{Diprotodon RGB image made of \ac{ASKAP} smoothed 15\arcsec\ (red), \ac{WISE} W3 (green) and \ac{WISE} W4 (blue). The dash box marks the nearby \HII\ region that is in the line of sight (Section~\ref{sec:IR}).}
    \label{fig:WISE}
\end{figure*}

In Figure~\ref{fig:WISE} we show a 4$^{\circ}\times4^{\circ}$ mosaic of the area centred on Diprotodon and its local environment as traced in the mid-infrared by the \ac{WISE} bands of W3 (12\,\um) and W4 (23\,\um)\footnote{We also examine other two \ac{WISE} bands, W1 (3.4\,\um) and W2 (4.6\,\um), but we did not use them in this study.}. The mosaics were created using the \ac{WISE} WXSC pipeline \citep{2012AJ....144...68J} with native resolution (6.5\arcsec\ in W3 and 12.0\arcsec\ in W4) and supersampled with 1\arcsec\ pixels. As described in \citet{2013AJ....145....6J,2019ApJS..245...25J}, the \ac{WISE} bands were carefully designed to sample both stellar emission (W1 and W2), and \ac{ISM} star-formation-excited gas/dust emission (W3 and W4). As expected, the stellar density is extremely high at these low Galactic latitudes, reaching the confusion limit of the \ac{WISE} imaging resolution for bands W1 and W2.  

Meanwhile, the star-formation-excited \ac{ISM} exhibits a complex filamentary structure, with tendrils reaching down from the Galactic Plane (bottom-right in Figure~\ref{fig:WISE}) penetrating through the \ac{SNR} shell (at least in projection). There are also signs of shock gas compression in the sharp filamentary structures to the west and northwest of the upper shell.
Similar to the \ac{SNR}, the \ac{ISM} emission is lower density on the eastern side, consistent with the radio-continuum (Figure~\ref{fig:1}) where the kinematics are hinting at a cavity blowout on the South and West side. Nevertheless, across this tangled \ac{ISM} emission is too complex to be directly associated with the Diprotodon or its initial blast wave.

\section{Results}
\label{sec:results}

We have serendipitously found a large-scale object in our new \ac{ASKAP}-\ac{EMU} survey \citep{2021PASA...38...46N} which we identify as a previously known Galactic \ac{SNR} \g\ (Figure~\ref{fig:1}). Following this \ac{ASKAP} Diprotodon detection, we also identified this radio source in the \ac{GLEAM} and \textit{Fermi}-LAT surveys. Our detection and classification of this Galactic \ac{SNR} is primarily based on the object morphology, size and multi-frequency appearance of Diprotodon with the method described in \citet[][Section 2.4]{2019PASA...36...48H}, \citet[][]{2022MNRAS.512..265F} and \citet{2023MNRAS.518.2574B}. 

\subsection{Diprotodon's Extent}
 \label{sec:radio_extent}

Previous measurements of Diprotodon's size and central position were limited by the instruments' poor resolution and sensitivity. It was suggested to have a diameter of $\sim$95\arcmin\ \citep{Green} as shown in Figure~\ref{fig:1} (green dashed circle). However, \citet{2020MNRAS.492.5980A} indicated that Diprotodon might extend in gamma-rays beyond the above estimate. Also, Fesen et al., (in prep.) showed new large-scale \Halpha\ and \OIII\ images of Diprotodon. The initial comparison with our radio-continuum images didn't show any feature alignment. This is no surprise as we have found similar tendencies in other specifically \OIII\ dominate \ac{SNR} as they are primarily confined to unusually low-density \ac{ISM} regions.
Here, we measured the extent of Diprotodon in all three: \ac{ASKAP} and \ac{GLEAM} radio-continuum and \textit{Fermi}-LAT gamma-ray images. 

We used the Minkowski tensor analysis tool BANANA\footnote{\url{https://github.com/ccollischon/banana}} \citep{2021A&A...653A..16C} to determine the centre and extent of Diprotodon. 
We found that Diprotodon \ac{SNR} is centred at RA(J2000)\,=\,9$^{\rm h}$59$^{\rm m}$50\fs5 and Dec(J2000)\,=\,--53\D19\arcmin50\arcsec\ ($l$=278\fdg94 and $b$=+1\fdg35). We measure Diprotodon to have an angular size of, at most, 200\arcmin$\times$194\arcmin\ (3\fdg33$\times$3\fdg23 at a position angle of 0\D). This makes Diprotodon one of the largest known \acp{SNR} in the sky according to the \citet{Green} catalogue and estimates by \citet{2019SerAJ.199...23S}, with the largest being the well-known Vela \ac{SNR} \citep[$\sim$8\D;][]{1998AJ....116.1886B}. Diprotodon is only somewhat smaller than the \acp{SNR} G70.0--21.5 (330\arcmin$\times$240\arcmin), G65.3+5.7 (310\arcmin$\times$240\arcmin), G74.0--8.5 (Cygnus Loop; 230\arcmin$\times$160\arcmin) and possibly G159.6+7.3 (240\arcmin$\times$180\arcmin). Finally, the recently discovered Galactic \ac{SNR} Hoinga \citep{2021A&A...648A..30B} also shows a large angular scale with a diameter of 4\fdg4 at a distance of $\sim$500\,pc.

Even more surprisingly, at a distance of 2.7\,kpc \citep{2019RAA....19...92S}, Diprotodon would have a staggering physical size of 157$\times$154\,pc, tucked in a presumably low-density region in between Carina-Sagittarius and Perseus spiral arm with its centre some 65\,pc above the Galactic Plane. From \citet{2019SerAJ.199...23S}, only Galactic \ac{SNR} G65.1+0.6 at a distance of 9.2\,kpc would have a larger diameter of 179.5\,pc. However, \citet{2020A&A...639A..72W} found that G65.1+0.6 is more likely to be closer at 4.16\,kpc, which makes its physical size $\sim$109\,pc ($\pm10\%$). \cite{2019SerAJ.199...23S} also list \ac{SNR} G312.4--0.4 at the distance of 14\,kpc --- to be of 154.8\,pc in size --- but again, optical extinction suggests a distance of only 4.4\,kpc, which makes its physical size significantly smaller. Therefore, if it has a distance of 2.7\,kpc, Diprotodon may be the physically largest Galactic \ac{SNR} known.

In the nearby \ac{MC}, we found a number of large-size \acp{SNR} and \ac{SNR} candidates \citep{2017ApJS..230....2B,2019A&A...631A.127M,2021MNRAS.500.2336Y,2023MNRAS.518.2574B,2024MNRAS.529.2443C}, some of which span up to 155\,pc. The two largest confirmed \ac{MC} \acp{SNR} are RXJ050736--6847.8 \citep[D=154.8\,pc; ][]{2021MNRAS.504..326M} and 0450--7050 \citep[D=120.2\,pc;][]{2009SerAJ.179...55C} would be of comparable size to Diprotodon. 

Certainly, there are several bubbles and superbubbles such as the \ac{LMC} 30\,Doradus~C that would be the same size as Diprotodon \citep{2021ApJ...918...36Y,2017ApJ...843...61S,2015A&A...573A..73K,2014AJ....147..162D}. However, they are all composed of multiple \ac{SN} explosions, and therefore, one could expect a larger extent. Another possible example of a large shell-like structure is the Galactic North Polar Spur with its suggested size of up to 300\,pc, which is likely a local superbubble wrapped just outside of our Local Bubble \citep{2021ApJ...923...58W}.

Such a large-size \ac{SNR} is expected to be rather aged ($\sim$10$^5$\,yrs) or in its late evolution stages. However, we note that Diprotodon's almost circular shape across such a large field of view suggests that the \ac{SNR} is still expanding in, presumably, rarefied ambient density.

\subsection{Diprotodon's Radio Spectral Index}
 \label{sec:radio_SED}

To estimate Diprotodon's radio spectral index (defined by $S \propto \nu^{\alpha}$, where $S$ is flux density, $\nu$ is the frequency and $\alpha$ is the spectral index), we used the \ac{GLEAM} detection in the 103--134\,MHz range (which gave the best signal-to-noise) and fitted a 2D plane to the background. We then integrated the flux density inside the \ac{SNR} defined polygon, where we avoided including bright sources in the background calculation. The total flux density of Diprotodon after background subtraction is 105$\pm$11\,Jy. 
As shown in the case of Ancora \citep{2023AJ....166..149F} and given Diprotodon's considerable angular size (especially for the low surface brightness and extended emission), we did not attempt to estimate Diprotodon's flux density at 943\,MHz.

\begin{figure}   
    \centering\includegraphics[width=\linewidth]{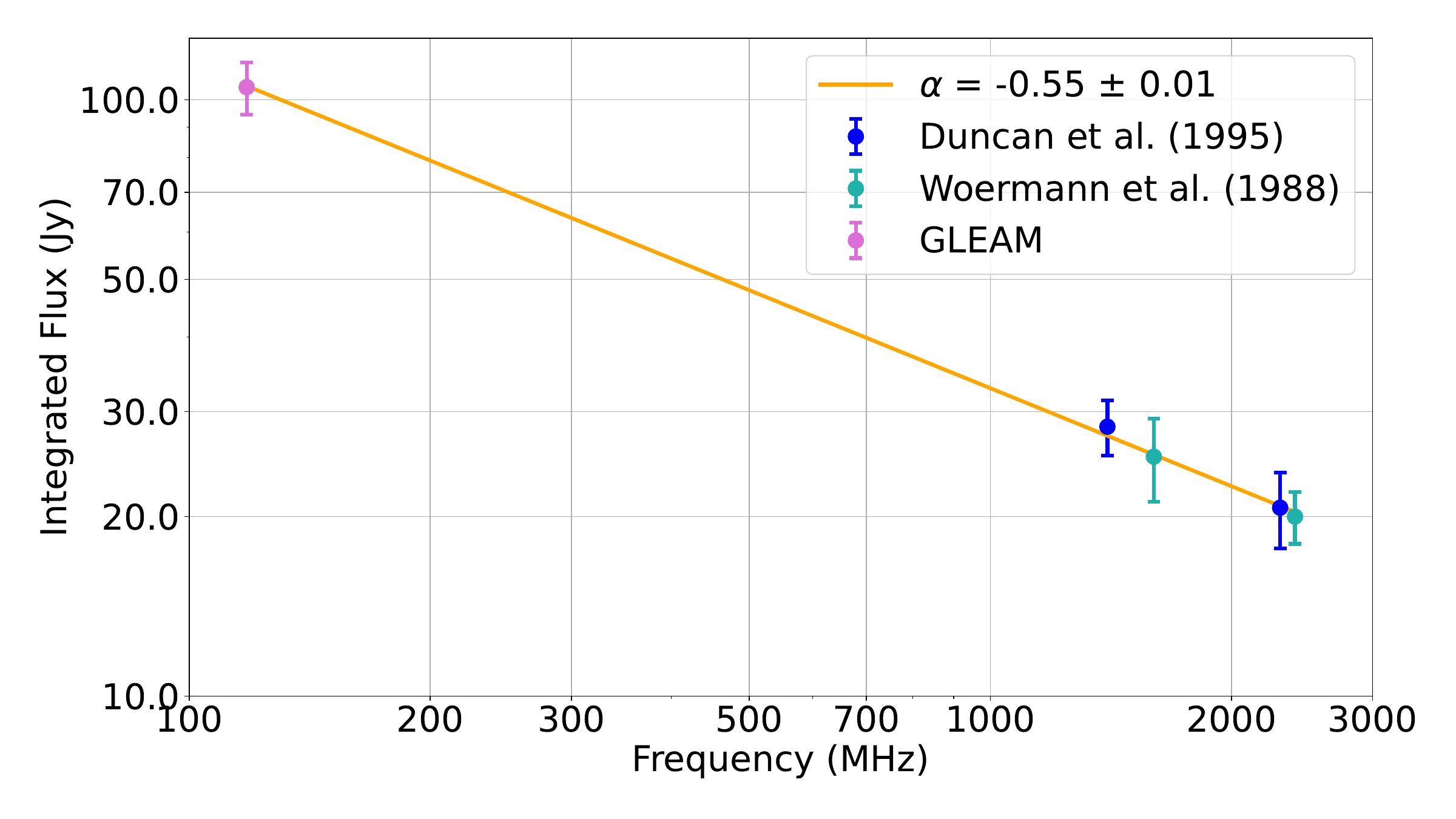}
    \caption{Diprotodon's radio flux densities as a function of frequency. A power-law fit based on flux density from \ac{GLEAM} (pink point), measurements from \citet{1988MNRAS.234..971W} (blue points) and  \citet{1995MNRAS.277..319D} (green points) with \ac{SNR} estimated $S_\mathrm{1\,GHz}$ = 32.8\,Jy and $\alpha = -0.55\pm0.01$.}
    \label{fig:sed}
\end{figure}

Using previous flux density measurements of \citet{1988MNRAS.234..971W} and \citet{1995MNRAS.277..319D} we obtain $\alpha=-0.55\pm0.01$ (Figure~\ref{fig:sed}). Combined with the above-measured angular size, this gives a radio surface brightness at 1~GHz of $\Sigma=1.3\times10^{-22}$~W~m$^{-2}$~Hz$^{-1}$~sr$^{-1}$ (assuming the emission to be spread smoothly over the \ac{SNR} boundaries defined above). However, we note that these earlier flux density measurements did not include the full extent of Diprotodon's radio continuum emission but, at the same time, included background source contribution. 

Nevertheless, this spectral index of $\alpha = -0.55\pm0.01$ is representative and typical for the average shell type \acp{SNR} observed for the Galaxy and range of nearby galaxies \acp{SNR} \citep{2012SSRv..166..231R, 2014SerAJ.189...15G, 2017ApJS..230....2B,2019A&A...631A.127M,book2,2023MNRAS.518.2574B,2023ApJS..265...53R}. It certainly strongly indicates that non-thermal radio emission dominates across Diprotodon apart from the western side, as also noted by \citet{1995MNRAS.277..319D}. This is where Diprotodon is superimposed with the most likely unrelated Galactic bubble E116 (centred at $l$=277\fdg725 and $b$=+0\fdg658) \citep{2019PASJ...71....6H} that can be seen in our WISE-IR images (Figure~\ref{fig:WISE}).

The above estimate of Diprotodon's radio surface brightness and originally suggested physical size of 157\,pc places this object well outside of the \ac{SNR} $\Sigma-D$ diagram \citep[][their fig.~3]{2020NatAs...4..910U,2018ApJ...852...84P}.
This indicates that either of the two estimates could be potentially wrong. 
One would expect that such a large \ac{SNR} would have a low surface brightness. Even if the estimate of Diprotodon's \ac{SNR} surface brightness is an order of magnitude lower ($\Sigma\sim10^{-23}$~W~m$^{-2}$~Hz$^{-1}$~sr$^{-1}$) it would still suggest an intrinsic diameter of $\sim$90-110\,pc. 
For a selected sample of 110 Galactic \acp{SNR}, \citet{2019SerAJ.199...23S} present a distance calibration for radio surface brightness at 1\,GHz vs diameter relation, using a cross-validation kernel smoothing. Diprotodon's 1\,GHz surface brightness of $\Sigma=1.3\times10^{-22}$~W~m$^{-2}$~Hz$^{-1}$~sr$^{-1}$ corresponds to the distance of $\sim$576\,pc and a diameter of $\sim$33\,pc. Given the significant amount of predicted missing flux density, this estimate should be taken as the lower limit of Diprotodon's distance and diameter.

\subsection{Diprotodon in H{\sc i} and CO Surveys}
 \label{sec:radio_HI}

To reveal the physical relation between Diprotodon and its surroundings, we analysed the archival \HI\ data taken from HI4PI \citep{2016A&A...594A.116H}, SGPS \citep{2005ApJS..158..178M} and the CO data from the NANTEN telescope \citep{2004ASPC..317...59M}.

\subsubsection{Diprotodon in the HI4PI and NANTEN Surveys}
\begin{figure*}
    \centering\includegraphics[width=\linewidth,clip]{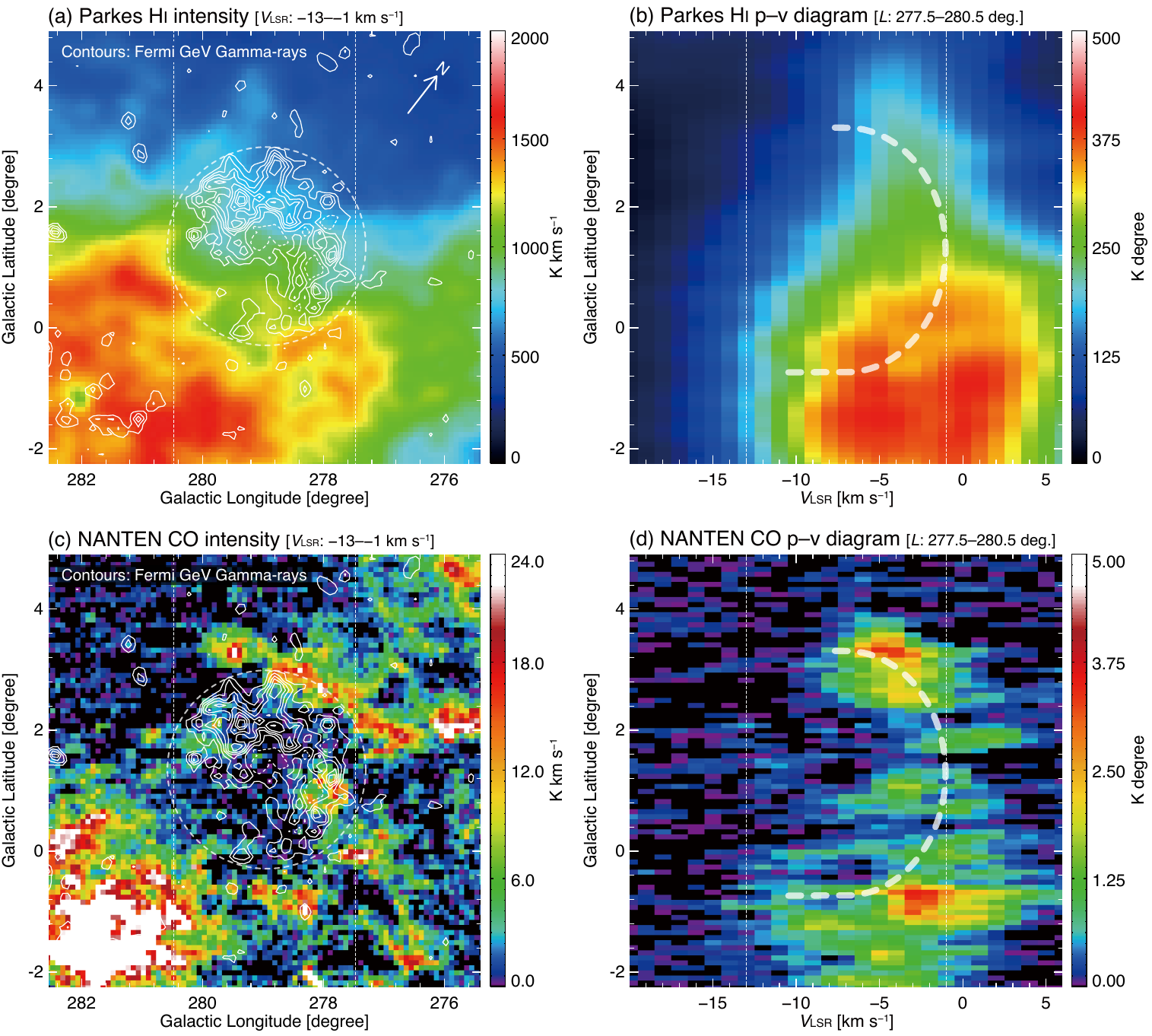}
    \caption{Velocity integrated intensity maps and position--velocity diagrams of \HI\ (a, b) and CO (c, d). The integration velocity range is from $-$13 to $-1$ km s$^{-1}$ for each intensity map and from 277\fdg5 to 280\fdg5 in Galactic latitude for each position--velocity diagram (dashed vertical lines). Superposed contours represent the GeV gamma-rays whose lowest contour and intervals are $5\sigma$ levels. The dashed white circle represents Diprotodon's radio shell boundary. The dashed curves in the position--velocity diagrams indicate the boundaries of the CO and \HI\ cavities (see the text).}
    \label{fig:HI}
\end{figure*}

Figures \ref{fig:HI}(a) and \ref{fig:HI}(c) show the distributions of \HI\ and CO superposed with the GeV gamma-ray contours and radio shell boundaries (dashed circles). We found a cavity-like distribution of \HI\ in the velocity range from $-13$ to --1\,\kms. The \HI\ clouds nicely trace the radio-shell boundary of Diprotodon, especially in the southern half of the shell. We also find filamentary molecular clouds in the northern edge of the \ac{SNR}. The other CO clouds are distributed as clumpy inside or on the edge of the radio-shell boundary. We note that there is no obvious correlation between the CO clumps and GeV gamma-ray peaks. Interestingly, such a spatial anticorrelation is also seen between the \HI\ and GeV-gamma-ray peaks. 

Figures~\ref{fig:HI}(b) and \ref{fig:HI}(d) show the position--velocity diagrams of \HI\ and CO. We find a hollowed distribution of CO and a velocity gradient of \HI\, whose spatial extents are roughly consistent with these of the radio-shell boundaries. Such spatial and velocity distributions of CO and \HI\ generally indicate an expanding motion of the molecular and atomic hydrogen gas, which could be formed by SN blastwaves and/or strong stellar winds from the progenitor system \citep[e.g.,][]{1990ApJ...364..178K, 1991ApJ...382..204K}. In other words, these \HI\ and CO clouds are possibly physically associated with Diprotodon. If this is the case, the systemic velocity of Diprotodon is in a range of Figure~\ref{fig:SEDs} (top-right panel), corresponding to an almost tangent velocity in this direction. Therefore, this implies that the smaller distance of $\sim$1.2\,kpc for Diprotodon is not in contradiction in terms of kinematic distances with other estimates shown in Sections~\ref{sec:SGPS} and \ref{sec:radio_age}. In any case, additional high-resolution \HI\ observations are needed to test this possibility.

To estimate the mass and density of molecular and atomic hydrogen gas, we used the following equations: 
\begin{eqnarray}
M_\mathrm{HI} = m_{\mathrm{p}} \Omega D^2 \sum_{i} N_i(\mathrm{H}{\textsc{i}}),\\
M_\mathrm{CO} = m_{\mathrm{p}}\Omega D^2 \sum_{i} N_i(\mathrm{H}_2),
\label{eq1_hi}
\end{eqnarray}
where $m_\mathrm{p}$ is the hydrogen mass, $\Omega$ is the solid angle for each data pixel, $D$ is the distance to the source, $N$(\HI) is the atomic hydrogen column density, and  $N$(H$_2$) is the molecular hydrogen column density. Here, we ignore the He abundance of molecular clouds. $N$(\HI) can be derived as $1.823 \times 10^{18}~W$(\HI), where $W$(\HI) is the \HI\ integrated intensity. We also used a relation $N$(H$_2$) = $2\times W$(CO) $\cdot$ $X_\mathrm{CO}$, where $W$(CO) is the integrated intensity of CO and $X_\mathrm{CO}$ is the CO-to-H$_2$ conversion factor. Here we adopted $X_\mathrm{CO} = 2\times 10^{20}$~cm$^{-2}$~(K~km~s$^{-1}$)$^{-1}$ \citep{2013ARA&A..51..207B}. We then obtained $M_\mathrm{HI} \sim 3.8\times10^5~M_{\odot}$ and $M_\mathrm{CO} \sim 3.3\times10^5~M_{\odot}$ within the radio shell boundary of the \ac{SNR}. The number densities of \HI, $n_\mathrm{p}$(\HI), and  H$_2$, $n_\mathrm{p}$(H$_2$), were also estimated to be $\sim$12~cm$^{-3}$ and $\sim$10~cm$^{-3}$, respectively. We then obtain the total interstellar proton density $n_\mathrm{p}$(H$_2$+\HI) to be $\sim$22~cm$^{-3}$. We note that this density represents the upper limit as it was calculated by summing up all gas densities along the line of sight.

\subsubsection{Diprotodon in the Southern Galactic Plane Survey}
 \label{sec:SGPS}

Using SGPS, we focus our attention on the long straight shell at the bottom of Diprotodon, which is parallel to the Galactic plane at a Galactic latitude of just below 0\D. To produce such a flat structure, the blast wave must be expanding in a strong density gradient in this direction. Such a density gradient is expected towards the Galactic plane. A density jump, which could be visible in the \HI\ line data, can also produce a flat straight shell such as the one we see in Diprotodon.

\begin{figure*}[!ht]   
    \centering\includegraphics[width=\linewidth]{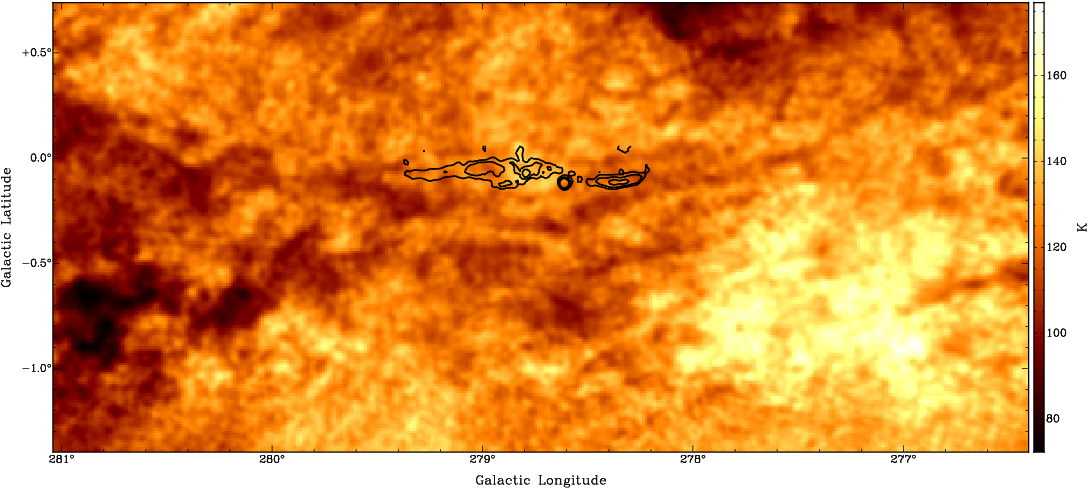}
    \caption{\HI\ line map taken from the SGPS \citep{2005ApJS..158..178M}. For this image, we averaged the three velocity channels between $-3.3$\,\kms\ and $-1.6$\,\kms. The black contours, taken from the continuum part of the SGPS \citep{Haverkorn2006}, indicate the flat southern shell of Diprotodon.}
    \label{fig:hisa}
\end{figure*}

\begin{figure}[t!]
    \centering\includegraphics[angle=-90, width=.9\linewidth]{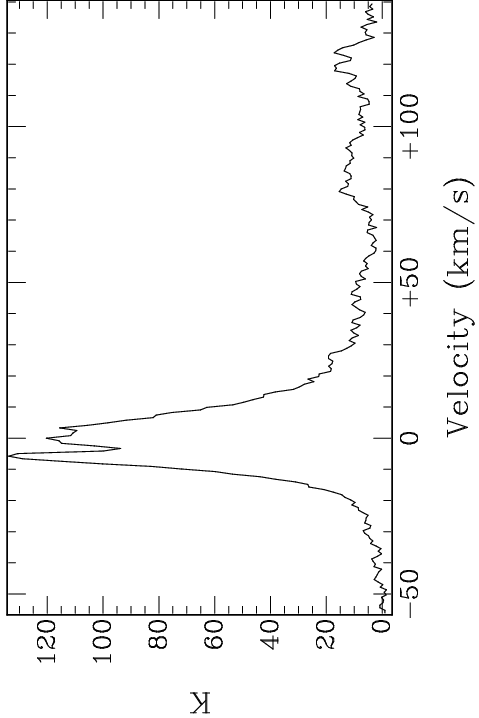}
    \centering\includegraphics[width=.9\linewidth]{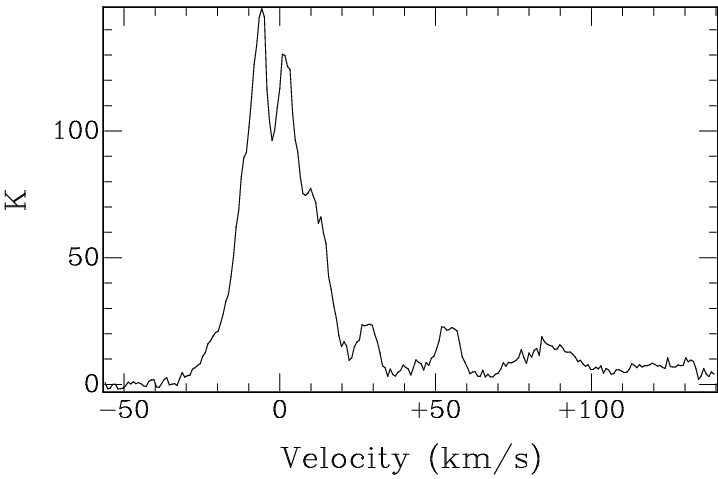}
    \centering\includegraphics[angle=-90,width=.9\linewidth]{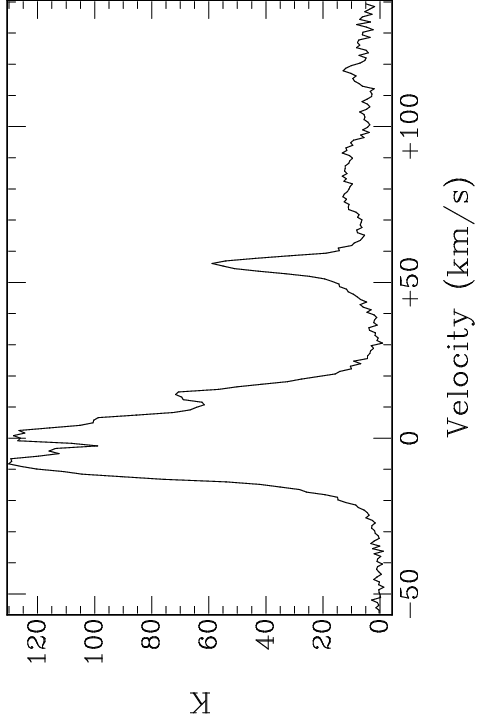}
   \caption{\HI\ spectra taken at three different positions just below the straight southern shell of Diprotodon. The spectra were taken at {279.33\D}, {$-0.18$\D} (top), {278.67\D}, {$-0.33$\D} (middle) and {278.41\D}, {$-0.23$\D} (bottom). Peak absorption velocities are $-3.3$~km\,s$^{-1}$ (top), $-2.5$~km\,s$^{-1}$ (middle), and $-2.6$~km\,s$^{-1}$ (bottom).}
    \label{fig:hisaspec}
\end{figure}

\begin{figure}  
    \centering\includegraphics[trim=0 0 80 385, width=\linewidth]{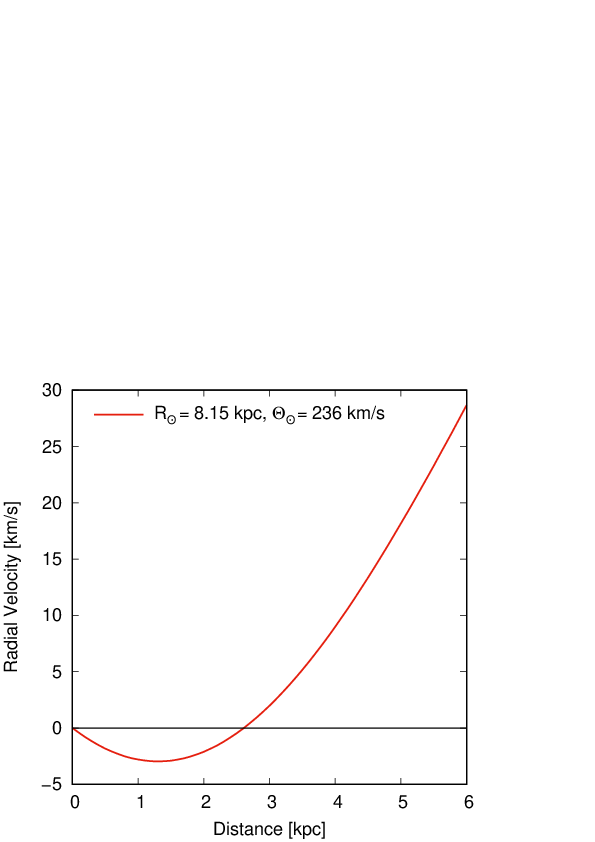}
    \caption{Galactic rotation curve in the direction of Diprotodon (Galactic Longitude = 279{\D}). We used the latest rotation curve published by \citet{Reid2019}, based on trigonometric parallax and proper motion measurements of molecular masers.}
    \label{fig:rotcurve}
\end{figure}

At small negative radial velocities, we find widespread dark absorption features in the \HI\ data from the SGPS indicative of \HI\ self-absorption (HISA) (see Figure~\ref{fig:hisa}). HISA is generated when Galactic rotation produces two different distances with the same radial velocity. If we find dense cold \HI\ gas at the near distance and warm \HI\ gas at the far distance, the bright emission of the background gas is absorbed by the denser foreground gas, producing dark shadows as seen in Figure~\ref{fig:hisa}. 
We find a concentration of HISA just below the bottom straight shell of the \ac{SNR}. The top border of the HISA is slightly curved, with Diprotodon sitting just above it. This is a strong coincidence indicating that Diprotodon's southern shell is running into dense \HI\ and possibly molecular gas, which is slowing down the expansion of the blastwave and forces the shell to follow the curved edge of the HISA area. 

In Figure~\ref{fig:hisaspec}, we show three sample \HI\ spectra taken just below the \ac{SNR}'s southern shell in the left, centre, and right parts. The narrow HISA absorption lines are obvious on top of the bright \HI\ background emission. They show centre velocities around $-3$\,\kms. 

We show the model rotation curve in the direction of Galactic longitude {279\D} in Figure~\ref{fig:rotcurve}. This is based on trigonometric parallax and proper motion measurements of molecular masers \citep{Reid2019}. We find negative radial velocities up to a distance of 2.5\,kpc. This is where the line of sight goes through the inner Galaxy with a tangent point velocity of about $-3$\,\kms\ at a distance of 1.2\,kpc. As a line of sight through the inner Galaxy naturally produces a rotation curve with two different distances at a given radial velocity, we find the perfect environment for the production of HISA. The direction we are looking at is along our own Galactic spiral arm. In the approximately opposite direction, there is the Cygnus~X region, which also exhibits lots of HISA \citep{Gottschalk2012}.

The association of Diprotodon with this HISA gives the \ac{SNR} a systemic velocity of about $-3$\,\kms. The radial velocity does not change much with distance (Figure~\ref{fig:rotcurve}), therefore the distance is not tightly constrained. However, as the \ac{SNR} is related to HISA, it must be at the near distance because, at the far distance, it cannot produce HISA. 
This puts an upper limit of 1.2\,kpc -- the location of the tangent point -- on the distance to Diprotodon which means that the \ac{SNR} is in our local spiral arm -- the Orion spur.

\subsection{Diprotodon's Location in the Milky Way}
 \label{sec:radio_age}
 
In Figure~\ref{fig:1}, which shows Diprotodon in galactic coordinates, we can see a clear flattening of the \ac{SNR} side that is touching $b$=0\D\, while the opposite side (at 2\D$<b<$3\D) is more circular. It is possible that we see the double shell morphology in the radio image. The lack of emission to the west could be due to the \HII\ region along the line of sight as shown in Figure~\ref{fig:1}, and not an intrinsic effect.

\begin{figure*}
    \centering\includegraphics[width=\linewidth]{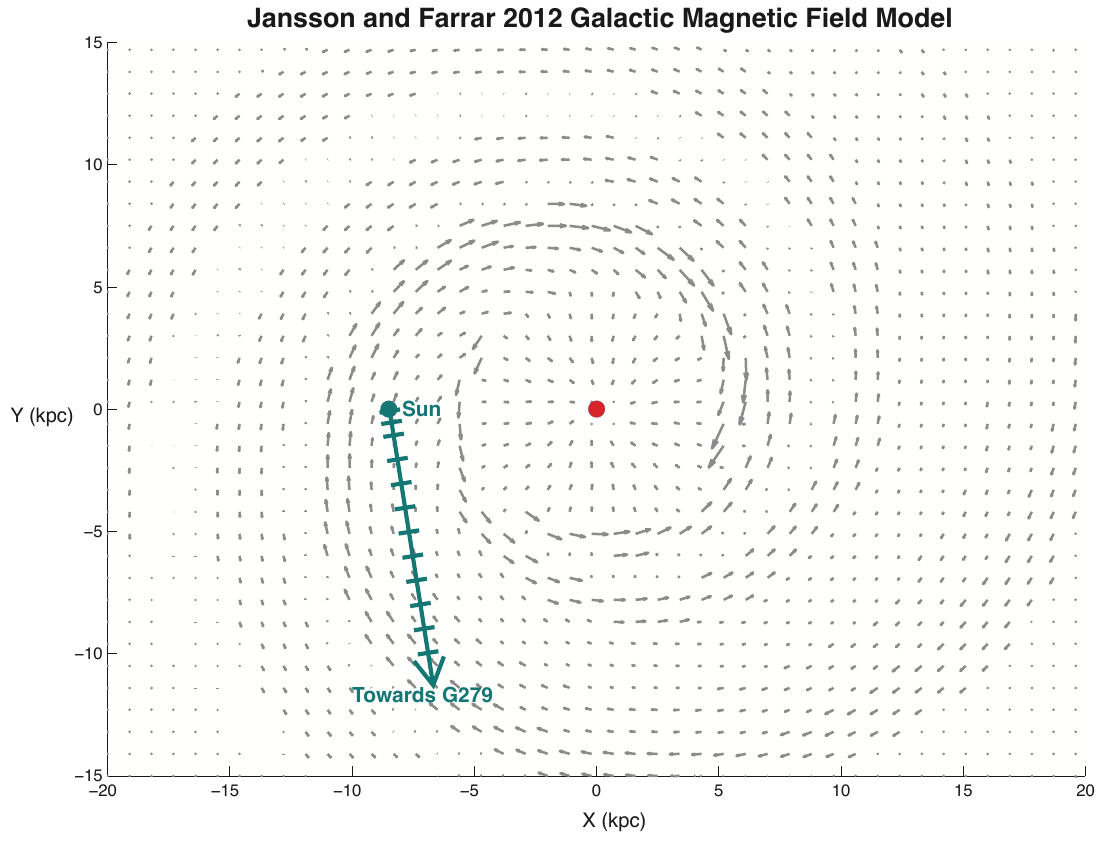}
    \centering\includegraphics[width=\linewidth]{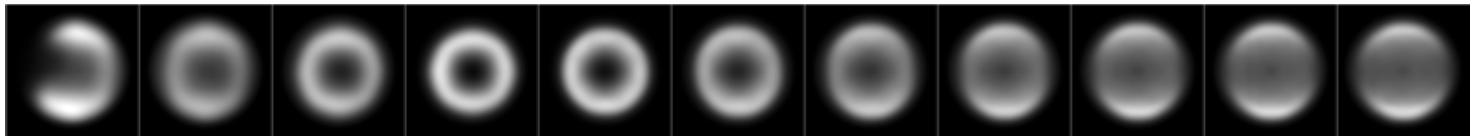}
    \caption{Top: Large-scale mean magnetic field model of JF12 shown in a top-down view. The arrow indicates the direction towards \g. The tick marks show 1~kpc steps along the line-of-sight, with the exception of the first tick, which is shown at 0.5~kpc. Bottom: Model images of the synchrotron emission that comes from an \ac{SNR} exploding into an ambient magnetic field defined by the JF12 model. The first model image is shown for a distance of 0.5\,kpc, and the subsequent model images show 1~kpc steps, from 1\,kpc to 10\,kpc (left to right).}
    \label{fig:GalBModel}
\end{figure*}

We examine the morphology of Diprotodon in the context of the large-scale Galactic magnetic field. Following the method of \citet{2016AA...587A.148W}, we model the synchrotron emission of an \ac{SNR} that explodes into the Galactic magnetic field model of \citet[][hereafter, JF12]{2012ApJ...761L..11J} towards the direction of Diprotodon. In the top part of Figure~\ref{fig:GalBModel}, we show a plot of the magnetic field vectors from JF12 with the direction to Diprotodon shown with an arrow. The bottom part of Figure~\ref{fig:GalBModel} shows the sequence of models shown at steps of increasing distance from 0.5, 1, 2, 3, etc. to 10\,kpc (left to right).

The models at the range of $\sim$2--5\,kpc distance look like a complete ring. 
This indicates that the model magnetic field for this direction and these distances is oriented nearly entirely along the line of sight.
The expected synchrotron intensity would be low with the magnetic field in this orientation. In addition, the geometry is also not consistent with the morphology that we observe in Diprotodon (see Figure~\ref{fig:1}). Comparing with Diprotodon's observed morphology, we find that the models are consistent with a distance $\leq 1$\,kpc or $>5$\,kpc, but since $>5$\,kpc would make Diprotodon's size unreasonably large, a distance $\leq 1$\,kpc seems the most likely scenario. 

Thus, this modelling does not support the 2.7\,kpc distance to Diprotodon. We suggest that it is consistent with \ac{SNR} being located much closer i.e. at $d_{\mbox{\tiny kpc}}\leq 1$\,kpc where $d_{\mbox{\tiny kpc}}$ is the distance in units of kpc. From now on, we consider that the distance to Diprotodon is at 1\,kpc, which gives it a dimension of 58$\times$56\,pc.

\subsection{Diprotodon's GeV Extension and Spectrum}
To measure the extent and location of the gamma-ray source, we performed a scan of the parameter space to maximize the likelihood for events with energies above 1\,GeV and a uniform disk hypothesis. The resulting size and location agree with those of the spatial template in the 4FGL-DR4 catalogue (4FGL~J1000.0--5312e), corresponding to an angular diameter of 2\fdg88, and we adopt this morphology for the rest of the analysis. 

We used events with energies above 0.2\,GeV to study the spectrum of the source. It is described by a log-parabola in the 4FGL-DR4 catalogue. We compared fits using a simple power law and a log-parabola, and the latter is preferred at the $6.6\sigma$ level when using the standard Galactic diffuse emission model ({\tt gll\_iem\_v07.fits}). However, we found that the spectrum below $\sim 1$\,GeV is significantly affected by the modelling of the diffuse Galactic emission. We repeated the analysis using the eight alternative models for the Galactic diffuse emission developed by \cite{2016ApJS..224....8A}\footnote{The alternative models were scaled appropriately to account for the differences in energy dispersion between {\tt Pass 7} and {\tt Pass 8} reprocessed data, see \url{https://fermi.gsfc.nasa.gov/ssc/data/access/lat/Model\_details/Pass8\_rescaled\_model.html}}. They were designed to explore some of the systematic effects on source parameters caused by uncertainties in the interstellar emission, including uncertainties in the cosmic ray source distribution, the height of the cosmic ray propagation halo and the spin temperature used to derive the H{\sc i} column density from the 21~cm line data. A simple power-law function was preferred for the Diprotodon spectrum in all the alternative models. A fit using a simple power law, $\frac{dN}{dE} = N_0 E^{-\Gamma}$, results in a spectral index $\Gamma = 1.680 \pm 0.005_{\mbox{\tiny stat}}\pm 0.14_{\mbox{\tiny sys}}$. We estimated the systematic error as in \cite{2016ApJS..224....8A} with the fits using the alternative diffuse emission models. More statistics at low energies will be needed to confirm or discard any spectral curvature for the \ac{SNR}. On the other hand, at higher energies, no spectral curvature is significantly detected above 1\,GeV using the standard Galactic diffuse emission model. The difference in the source TS in fits using a log-parabola and a simple power law is 5.1. The spectral index and luminosity above 1\,GeV are $1.79\pm 0.02_{\mbox{\tiny stat}}$ and $\sim 5.8\times10^{33}\,d_{\mbox{\tiny kpc}}^2$~erg~s$^{-1}$, respectively.

To get flux points, we divided the data into 15 logarithmically-spaced energy intervals and fit the spectral normalisation of 4FGL~J1000.0--5312e, assuming a simple power law spectrum with index 2 in each interval. We estimated a systematic uncertainty of 13\%\ for the spectral normalisation in the 0.2--500\,GeV energy range using the alternative diffuse emission models described above and added this uncertainty to the statistical uncertainties of the flux points in quadrature.


\section{Diprotodon's Mystery}
 \label{sec:mistery}

 \subsection{Diprotodon's Evolutionary Status}
 \label{sec:evolution}

Diprotodon's multi-frequency filamentary structure (morphology) and spectral index are strong indications of predominantly synchrotron emission from electrons. Therefore, we should expect gamma rays from the same particles. The fact is that \citet{2011MNRAS.414.2282S} detected enhanced \SII\ emission across several Diprotodon's filaments, suggesting that they are still in the radiative phase. It is most likely that Diprotodon grew to such an extent in a low-density environment so it can still have a well-defined forward shock.

To estimate and evaluate the true evolutionary status of Diprotodon, we use the above-mentioned $\Sigma-D$ evolutionary tracks and a more realistic diameter estimate of $\sim$58\,pc (Section~\ref{sec:radio_age}). 
As Diprotodon is positioned at the bottom right corner of the $\Sigma-D$ diagram, we can conclude that it is an evolutionary advanced and low surface brightness \ac{SNR} in the radiative phase of evolution, and in this way is similar to the \ac{SNR} Ancora \citep[see:][their fig.~4]{2023AJ....166..149F,2024A&A...684A.150B}. It most likely evolved in a medium-density interstellar environment with densities $\sim$0.2\,cm$^{-3}$ and the energy of the \ac{SN} explosion can be assumed to be the canonical 10$^{51}$\,erg. 

The next important issue for the determination of Diprotodon's evolutionary stage is from associated magnetic field values. 
To estimate Diprotodon's magnetic field, we used the equipartition model from \url{http://poincare.matf.bg.ac.rs/~arbo/eqp}. This method uses modelling and simple parameters to estimate intrinsic magnetic field strength, energy contained in the magnetic field and cosmic ray particles using radio synchrotron emission \citep{2012ApJ...746...79A,2013ApJ...777...31A,2018ApJ...855...59U}.
We use: $\alpha=-0.55$, radius $\theta$~=~98\,arcmin, $\kappa=0$, $S_{\rm 1\,GHz}$~=~32.8\,Jy, and $f=0.018$ and found that the mean electron equipartition field over the whole of Diprotodon varies from $13.7\,\mu$G for 2.7\,kpc distance to $19.6\,\mu$G for distance of 0.75\,kpc distance, with an estimated minimum energy of $E_{\rm min}=1.7\times10^{49}-7.6\times10^{47}$\,erg. 
The original model, developed in \citet{2012ApJ...746...79A}, yields a mean ion equipartition ($\kappa\neq 0$) field of 35.7\,$\mu$G (for distance 2.7\,kpc) to 51.2\,$\mu$G (for distance 0.75\,kpc), with an estimated minimum energy of $E_{\rm min}=1.2\times10^{50}-5.2\times10^{48}$\,erg.

\subsubsection{Diprotodon in the radiative phase}

Even at a distance of $\sim$1\,kpc and diameter of $\sim$57\,pc, Diprotodon is in the radiative phase and not the Sedov-Taylor or adiabatic phase. To determine the evolutionary phase of an \ac{SNR}, we can compare the width of the shells and filaments with the radius of the remnant. For a Sedov-Taylor phase \ac{SNR} the compression ratio is 4 \citep{Sedov1959}, which translates into a shell width-to-radius ratio of about 1:10. A younger \ac{SNR} would have a much wider shell, as we can detect synchrotron emission from the outer shell and the convection zone between swept up material and ejecta \citep[e.g.][]{Gull1973}. For a radiative \acp{SNR}, we find large compression ratios of several 10s or even 100s. 

To determine the width of an \ac{SNR}’s shell we have to consider that this is a three-dimensional object and the emission we observe is projected onto the plane of the sky. We cannot simply measure the width of the projected shell, as it also contains emission from the spherically expanding shell part that is moving away from us and towards us. The emission peaks of the shells mark the longest line of sight through the synchrotron emitting region. This is the inner edge of the shell projected onto the plane of the sky. The radial distance between that peak and the outer edge of the shell emission indicates the width of the actual three-dimensional shell. In Diprotodon, which shows a wealth of thin filaments that are expected for a fragmenting radiative forward shock, this width is typically not greater than 1\arcmin. The projected emission from the fragmented shell fills a large projected volume in the northeast, compared to the south, where only one isolated filament is present. This gives a ratio of the shell width to the radius of the \ac{SNR} of around 0.006, which translates to a compression ratio of about 50, which is clearly well beyond the Sedov-Taylor phase. A more quantitative analysis would require the extraction of radial emission profiles, which is beyond the scope of this paper based on the available data.

Using the study of radiative \acp{SNR} by \citet{Cioffi1988}, we can calculate the time $t_{\rm PDS}$ and radius $R_{\rm PDS}$ at which an \ac{SNR} enters the radiative phase, or as they call it the pressure driven snowplow (PDS) phase, given intrinsic parameters for the explosion energy, $E_{51}$ given in $10^{51}$~erg, and the ambient number density, $n_0$ given in cm$^{-3}$:
\begin{equation}
    t_{\rm PDS} = 1.33\times 10^4 \frac{E_{51}^{3/14}}{n_0^{4/7}}~{\rm yr}
\end{equation}
\begin{equation}
    R_{\rm PDS} = 14.0 \frac{E_{51}^{2/7}}{n_0^{3/7}}~{\rm pc}.
\end{equation}

Assuming a distance of $d_{\mbox{\tiny kpc}}$=1\,kpc gives Diprotodon an average diameter of $\sim$60\,pc. For explosion energies $E_{51}$ of 0.1 and 1.0, which can be considered lower and upper limits, we derive ambient densities $n_0$ of 0.17\,cm$^{-3}$ and 0.036\,cm$^{-3}$ and ages $t$ of 54\,kyr and 36\,kyr, respectively, for the time the \ac{SNR} enters the PDS phase. As Diprotodon is clearly well into the radiative phase, the values for the ambient density and the ages are lower limits.

\begin{table}[]
    \centering
    \begin{tabular}{|lr|ccccc|} \hline
     & $n_0$ & 0.05 & 0.10 & 0.20 & 0.50 & 1.0  \\ 
    $E_0$ & & & & & & \\ \hline
     \multicolumn{2}{|l|}{1.0} & S-T & S-T & $4.0\times 10^4$ & $7.6\times 10^4$ & $9.8\times 10^4$ \\
      \multicolumn{2}{|l|}{0.32} & S-T & $5.1\times 10^4$ & $8.0\times 10^4$ & $1.7\times 10^5$ & $3.0\times 10^5$ \\
      \multicolumn{2}{|l|}{0.1} & $6.4\times 10^4$ & $1.0\times 10^5$ & $1.7\times 10^5$ & $6.8\times 10^5$ & merged \\ \hline
    \end{tabular}
    \caption{Age $t$ in yr for Diprotodon for different intrinsic parameters $n_0$ in cm$^{-3}$ and $E_0$ in $10^{51}$\,erg. The distance was assumed to be 1\,kpc, which translates to a mean radius of about 30\,pc. ``S-T'' indicates that the \ac{SNR} would not have reached the PDS phase yet. ``Merged'' indicates that this \ac{SNR} would merge with its environment before reaching a radius of 30\,pc.}
    \label{tab:radpar}
\end{table}

In Table~\ref{tab:radpar}, we show some possible scenarios for Diprotodon for three different explosion energies between a lower limit of $10^{50}$\,erg and an upper limit of $10^{51}$\,erg for a Type-Ia explosion. The $3.2\times 10^{50}$~erg are the logarithmic average of the lower and upper limit and correspond according to \citet{Pejcha2015} to a type~IIP explosion with an ejecta mass of 10~$M_\odot$. If we now assume $n_0 = 0.2$\,cm$^{-3}$ and $E_0 = 10^{51}$\,erg, suggested as a possible scenario based on the radio surface brightness study in Section~\ref{sec:evolution}, Diprotodon would be about 40\,kyr. 

Another possible scenario is a well-structured environment in which the \ac{SNR} first expands inside a low-density cavity and enters the radiative phase by crushing into higher-density clouds. This may explain the higher-than-expected radio surface brightness for such an extensive \ac{SNR} and the bright gamma-ray emission, as the evolutionary development of the \ac{SNR} is stalled until it reaches the clouds. However, in this scenario, we would expect to find a cloud outside of each bright, highly compressed radio shell, which is clearly not seen in our data.

Here argued the radiative phase of Diprotodon's evolution is at odds with the eROSITA X-ray study \citep{2024A&A...685A..23M} that shows an oxygen-rich, ejecta-dominated \ac{SNR}, with the plasma in a non-equilibrium state, favouring a somewhat younger age and a smaller size of only $\sim$20\,pc. 

While temperatures of 0.3 or 0.6\,keV point to cooler and more evolved remnants, but that could be due to interaction with denser material that agrees with the gamma-ray emission.
Nevertheless, it is important to keep in mind that not all properties of this remnant necessarily indicate that it is among the largest or oldest.

\subsection{Large and Aged SNR That Shines in Gamma-Rays}
 \label{sec:oldsnr}

The most unusual fact is that such an extensive \ac{SNR} is detected in GeV gamma-rays with a hard spectrum. This is certainly quite unexpected and challenges our understanding of the evolution of such objects. This type of gamma-ray spectrum is typical of young (and therefore smaller) TeV shell-like \acp{SNR} such as RCW~86, Vela~Jr and RX~J$1713.7-3946$ \citep{2016ApJ...819...98A,2011ApJ...740L..51T,2011ApJ...734...28A}. An example of a puzzling \ac{SNR} with an angular size of $\sim$3\D\ that shows a dynamically young GeV spectrum while exhibiting morphological features of an evolved object is G150.3+4.5 \citep{2014A&A...567A..59G,2020A&A...643A..28D}. There are other objects seen in gamma rays that could be (evolved) \acp{SNR}, such as the shell-like TeV source HESS\,J1912+101 \citep{2008A&A...484..435A,2019RAA....19...45R}; and others of unknown type with hard GeV spectra having no known counterparts at lower energies such as G350.6--4.7 \citep[3\fdg4 wide,][]{2018MNRAS.474..102A,2018ApJS..237...32A} and 2HWC\,J2006+341 \citep{2020ApJ...903L..14A}. However, none of the known evolved Galactic \acp{SNR} have a hard GeV emission while also having a similar physical size to Diprotodon.

There are a few aspects to this problem. So far, the sample of detected gamma-ray sources is somewhat biased to what we can actually see. Young ($\sim$1000~years) \acp{SNR} are predominantly leptonic as they usually evolve in low-density media, but can efficiently accelerate electrons. With time, leptonic emission (inverse-Compton) becomes suppressed due to electrons losing energy via synchrotron radiation. The TeV emission starts to decrease after $\sim$3000~years \citep{2020A&A...634A..59B}, but GeV emission should also start to decrease at some point. So, for aged \acp{SNR}, we should not see much gamma-ray emission from electrons. On the other hand, dynamically old \acp{SNR} that we observe are predominantly hadronic with a soft spectrum \citep{2013Sci...339..807A,2011MmSAI..82..747G,2016ApJ...816..100J,2019A&A...623A..86A, 2020MNRAS.497.3581D}, which agrees with theoretical expectations and can be explained as a combined effect of the decrease of the maximum energy to which particles can be accelerated with time and particle escape \citep{2019MNRAS.490.4317C, 2020A&A...634A..59B}, or alternatively by weakening of the shock due to propagation in the hot shocked wind of the progenitor star \citep{2022A&A...661A.128D}. 
Protons do not lose energy as electrons and hence can generate gamma-ray emission at late times. Dynamically old \acp{SNR} also usually show strong evidence of cloud interaction that causes deceleration of the shock and provides target material for hadronic interaction.

Supposedly, the age ($\sim10^4-10^5$ years) of Diprotodon naturally indicates a hadronic nature of the GeV gamma-ray emission. However, this scenario is in tension with a lack of clear correlation of the gamma-ray emission with gas distribution (see Section~\ref{sec:radio_HI}). The hadronic scenario also requires protons to be accelerated to a hard spectrum up to $\sim10$~TeV, which would be somewhat surprising (see further discussion in Section~\ref{sec:SED}). The observed hard gamma-ray spectrum could be more naturally explained in the case of a much younger age of the remnant both in the hadronic scenario and the leptonic scenario (through the curvature of the IC spectrum imposed by the cut-off in the electron spectrum and/or the spectral break resulted from synchrotron cooling).

\subsection{Diprotodon's Spectral Energy Distribution Modeling}
 \label{sec:SED}

To help understand the particle nature of the {\em Fermi}-LAT emission from Diprotodon, we applied a model that matches the non-thermal emission from a steady-state population of particles (electrons and protons) accelerated by Diprotodon in a single zone of magnetic field $B$ to the radio and GeV fluxes. The model assumes particles are injected according to a power law + exponential cutoff distribution $\sim E^{-\Gamma}\exp(-E/E_{\rm cut})$ with photon index $\Gamma = 2$. The exponential cutoff energy $E_{\rm cut}$ is used to represent the maximum particle energies ($E_{\rm cut,e}$ and $E_{\rm cut,p}$ for electrons and protons respectively)  expected from acceleration limits, radiative energy losses, and potentially particle losses due to their escape from the \ac{SNR} region. The total energy of protons and electrons (erg) is denoted $W_p$ and $W_e$, respectively. 

Photon production from synchrotron, inverse-Compton (IC), Bremsstrahlung, proton-proton (PP) collisions and secondary synchrotron processes were considered. The PP collision cross-section and secondary particle distributions are defined by \citet{Kafexhiu:2014}. The IC emission is assumed to result from three low-energy photon fields -- the cosmic microwave background (CMB), infrared photons, and optical photons from starlight. The PP collision rate is governed by the interstellar medium target density $n$ (cm$^{-3}$), where we used $n=20$\,cm$^{-3}$ ($n=5$\,cm$^{-3}$ for the leptonic dominated case) based on our \ac{ISM} studies discussed earlier. We note that this value represents an upper limit calculated from the column density along the line of sight.  

We assume Diprotodon to be at a 1.0\,kpc distance (see Section~\ref{sec:radio_HI}), resulting in a physical diameter of $D_{\rm SNR}$=58\,pc. 
We consider contributions from both the leptonic and hadronic components to the {\em Fermi}-LAT GeV emission.
For the IC seed photons, we used values typical of the Galactic average for the infrared and optical photon fields \citep[e.g. see][]{Vernetto:2016}, as the infrared image (Figure~\ref{fig:WISE}) and O-star catalogues do not suggest any significant enhancements. Thus, we assume energy densities of 0.3\,eV\,cm$^{-3}$ for both of these fields and thermal black-body temperatures of 30\,K and 2600\,K, respectively, for the infrared and optical fields.

Some constraints are also available from the X-ray and TeV gamma-ray domains. Just recently, \citet{2024A&A...685A..23M} revealed a thermal X-ray counterpart to the \ac{SNR} using observations with the eROSITA telescope. They also noted some hints for this emission in their analysis of ROSAT PSPC data. The thermal X-ray spectra are consistent with a dual-temperature shock-heated gas model (with a number of spectral lines), with $kT$=0.34 and 0.60\,keV respectively. In our model, we treat this as an upper limit to any non-thermal X-ray emission by including a dual-temperature thermal Bremmsstralung approximation ($\sim T^{-0.5}\exp(-E/kT)$) normalised to the total energy flux, 1.48$\times 10^{-9}$\,erg\,cm$^{-2}$\,s$^{-1}$ quoted by \citet{2024A&A...685A..23M} in the 0.2 to 4.0\,keV energy range.     
For upper limits in the TeV gamma-ray range, we used the available integral $E>1$\,TeV flux upper limits from the H.E.S.S. Galactic Plane Survey (HGPS) \citep{HGPS} summed over a 0\fdg2 region. We first found the 0\fdg2 upper limit averaged over the \ac{SNR} region and then scaled this value by the ratio $R_{\rm SNR}$/0\fdg2. This result was then differentiated and converted to an energy flux in the 1 to 10\,TeV energy range assuming a differential photon index $\Gamma=2.3$ as assumed by \citet{HGPS}. 

\begin{figure*}[t!]
\centering
\hbox{
\includegraphics[width=0.5\textwidth]{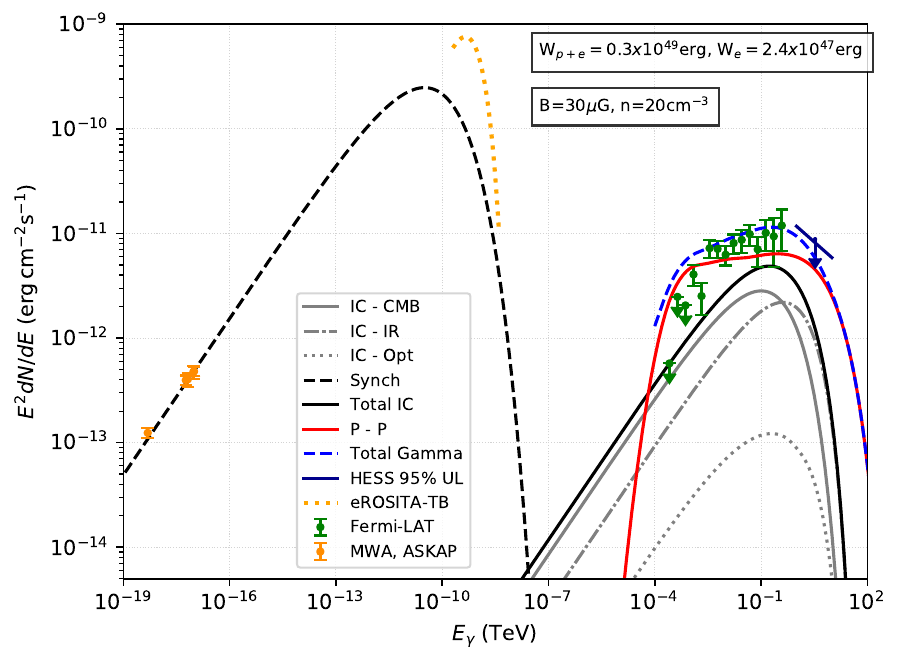}
\includegraphics[width=0.5\textwidth]{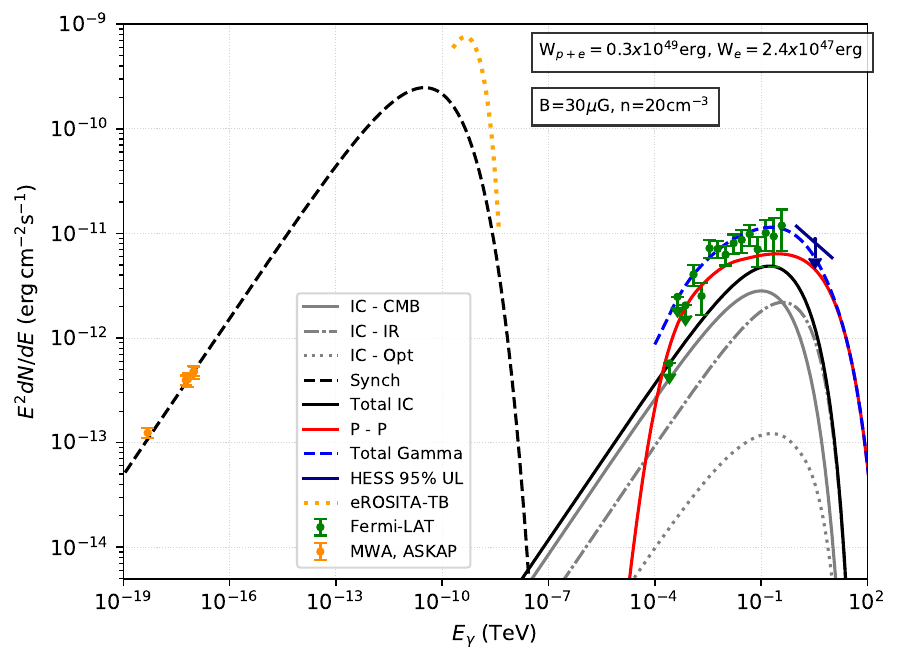}
}

\hbox{
\includegraphics[width=0.5\textwidth]{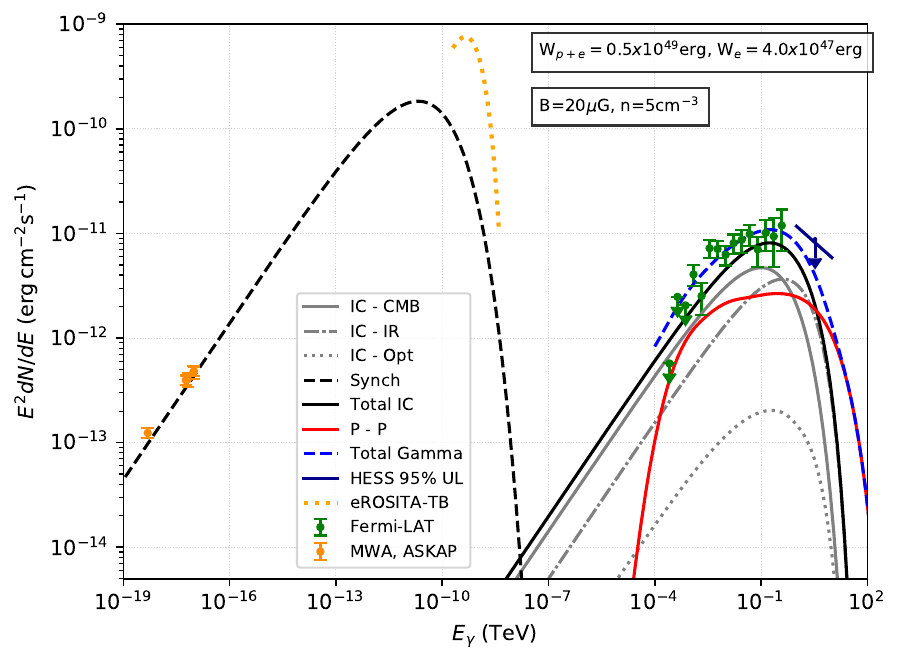}
\includegraphics[width=0.5\textwidth]{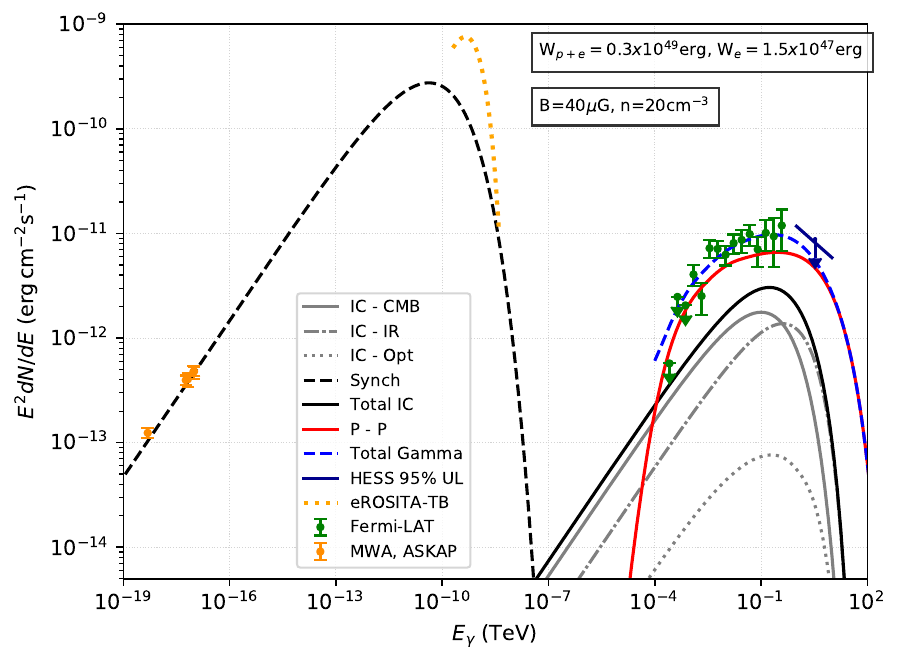}
}
\caption{Top-left: Model spectral energy distribution from hadronic (PP) and leptonic (synchrotron and IC emission) particle populations applied to Diprotodon for the \ac{SNR} at 1.0\,kpc distance with radius $R_{\rm SNR}=29$\,pc. Cutoff energies $E_{\rm cut, e,p}$ = 5 and 100\,TeV were used for the injected electrons and protons. Additional model parameters are shown. The top-right panel uses a modified density $n^\prime(E_p)$ in the PP component that considers energy-dependant penetration into dense \ac{ISM} clumps over an age 35\,kyr. A clump radius $R_{\rm cl}=0.1$\,pc and clump magnetic field $B_{\rm cl} = 6B$ were assumed. The bottom-left and bottom-right panels illustrate minor variations in magnetic field, $W_p$ and $W_e$ to produce leptonic- and hadronic-dominant scenarios, using the same particle cutoff energies as per the top panels, and a modified density as per the top-right panel. The 95\% confidence level HESS upper limit is shown on all panels (converted from an integral limit for $E>1$\,TeV to a differential limit over the 1 to 10\,TeV energy range assuming a spectral index of $-$2.3 as used by \citet{HGPS}). In all panels, the eROSITA thermal Bremsstrahlung component (eROSITA-TB, see text) is treated as an upper limit to the synchrotron emission.} 
\label{fig:SEDs}
\end{figure*}

Results are shown in Figure~\ref{fig:SEDs} along with the parameters found to adequately match the radio and {\em Fermi}-LAT fluxes. A reasonable match was found when assuming a magnetic field $B=30\mu$G, $E_{\rm cut,p}=100$\,TeV, $E_{\rm cut,e}=5$\,TeV, total energy $W_{p+e}=0.3\times10^{49}(d/1\,{\rm kpc})$\,erg and electron energy assumed to be $W_e=2.4 \times 10^{47}(d/1\,{\rm kpc})$\,erg. These particle energy budgets are similar to the canonical expectation for an \ac{SNR} and to local measurements of the electron spectrum. Emission components from the non-thermal Bremsstrahlung and secondary synchrotron processes are negligible and are therefore not shown. 
The 5\,TeV electron cutoff energy is constrained by the eROSITA thermal emission and is consistent with the expectation for mature ($>10^4$\,yr) \acp{SNR}, where maximum electron energies are limited by synchrotron losses (e.g. \citet{Reynolds:2008}). The 100\,TeV proton cutoff energy as constrained by the TeV upper limit from H.E.S.S. represents the upper limit on the maximum proton energy. The GeV emission up to $\sim$400\,GeV is clearly positioned inside the radio structures, implying that particles up to at least $\sim$4\,TeV or so are confined at the current epoch. Although it is expected that an \ac{SNR} could accelerate protons up to 100~TeV and higher, these high energies are normally reached only during the early stages of the \ac{SNR} evolution \citep[e.g.][]{2013MNRAS.431..415B}. The hard broad-band spectrum of the GeV gamma-ray emission suggests the presence of recently accelerated protons to at least 4\,TeV. However, such proton acceleration in mature SNRs is not expected to reach significantly higher energies than this due to 
deceleration of the shock and less effective amplification of the magnetic field (e.g. \citet{2020A&A...634A..59B}). We do note that a lower proton cutoff energy approaching 10 to 20\,TeV in our model would also provide a satisfactory match to the data if we were not aiming to meet the H.E.S.S. upper limit.

For energies below a few GeV, we note that our PP model overestimates the observed fluxes and upper limits by factors of about 2 to 5 (Figure~\ref{fig:SEDs} top-left panel). In the framework of our one-zone stationary model, this could be addressed by using a slightly harder proton spectrum, e.g. $\Gamma$=1.8 to 1.9. However, the radio fluxes from the \ac{SNR} suggest $\Gamma \sim 2.0$ to 2.1 for electrons, and recent theoretical studies also support the likelihood of indices steeper than 2.0 in \ac{SNR} shock acceleration (e.g. \citet{Bell:2019,Malkov:2019}), although the potential for harder indices in shock acceleration for some circumstances has been discussed (e.g.\citet{Malkov:1999,Perri:2012}). The vast majority of dynamically old hadronic dominated GeV gamma-ray \acp{SNR} exhibit a very soft spectrum \citep{2013Sci...339..807A,2011MmSAI..82..747G,2016ApJ...816..100J,2019A&A...623A..86A, 2020MNRAS.497.3581D}. However, along with the Diprotodon \ac{SNR} we study here, a few other SNRs (presumably mature in age) with a hard spectrum below 10\,GeV have also been detected (e.g. \citet{Zeng:2021,2022MNRAS.510.2920A,Eppens:2024}).

A potential explanation for the hard GeV spectrum, if viewed in terms of a hadronic scenario, might come from the energy-dependant diffusive penetration of protons into the \ac{ISM} gas, which could be important in the case of clumpy \ac{ISM} inside the \ac{SNR} if it results from a core-collapse supernova event. In this case, the dense (10$^{3\, {\rm to}\, 4}$\,cm$^{-3}$) clumps that survive the \ac{SNR} shock and gas turbulence induced by it may comprise a reasonable fraction of the gas mass downstream of the shock. This issue has already been discussed in application to the young (1600\,yr), gamma-ray-bright \ac{SNR} RXJ\,1713.7$-$3946, where a very hard ($<100$\,GeV) spectrum is observed (e.g. \citet{Inoue:2012,Gabici:2014}), and in the GeV emission from nearby molecular clouds \citep{Yang:2023}. Recent measurements of the \ac{ISM} towards some \acp{SNR} with ALMA have revealed the presence of sub-pc-scale \ac{ISM} clumps (e.g. \citet{SanoRXJ713:2020,SanoW49B:2021}). Following \citet{Inoue:2012} the penetration depth of protons in pc units can be given by $l_{\rm pd}(E_p) = 0.1 \sqrt{\eta (E_p/10{\rm TeV}) (t_{\rm age}/10^3{\rm yr}) / (B_{\rm cl}/100\,\mu{\rm G})}$. Here, $E_p$ is the proton energy, $t_{\rm age}$ is the \ac{SNR} age, $B_{\rm cl}$ is the clump magnetic field and $\eta\sim (B/\delta B)^2$ is a turbulence parameter based on magnetic field fluctuation estimates. 

As a potential first look at this effect as applied to our model, we follow \citet{Inoue:2012} who suggest $\eta=1$ and a clump radius of $R_{\rm cl} = 0.1$\,pc. To adjust our model PP component, we apply an effective \ac{ISM} density $n^\prime(E_p) = n \zeta(E_p)$, where $\zeta(E_p) = 1 - [(R_{\rm cl} - l_{\rm pd}(E_p))/R_{\rm cl}]^3$ represents the energy-dependent ratio of clump volume intercepted by protons to the total clump volume, under a constant clump density assumption. We note that $\zeta(E_p)$ is upper bound to 1.0 for cases where the protons fully penetrate the clump. Detailed simulations of the \ac{SNR} shock influence on dense clumps (e.g. \citet{Inoue:2012,Celli:2019}) suggest clump magnetic fields can reach values of 100$\mu$G  or more inside and around the shock-disrupted boundary. To achieve these values, in our application, we simply assume the clump magnetic field is 6$\times$ larger than the \ac{SNR} averaged field, $B_{\rm cl} = 6 B$. Results in Figure~\ref{fig:SEDs} (top-right panel) assuming an age 35\,kyr show that with the modified density, the PP component $<10$\,GeV can be satisfactorily reduced to match the observations.

Our simple model assumes that (1) most of the gas is in the form of clumps and (2) the gamma-ray emissivity dominantly arises from these clumps. For the gas density averaged over the \ac{SNR} volume $n= f n_{\rm cl} + (1-f)n_{\rm icl}$ for a clump volume filling factor $f$, clump density $n_{\rm cl}$  and inter-clump density $n_{\rm icl}$, condition (1) requires $n_{\rm cl} >> (1-f) f^{-1} n_{\rm icl}$. Following this, from condition (2), it can be shown that we require $n_{\rm cl} >> (1/\zeta(E_p) -1) n_{\rm icl}$. A further condition, that the clumps do not significantly affect the \ac{SNR}'s dynamical evolution (compared to effects produced by large pc-scale gas clouds), would require a small filling factor $f<$0.1 (e.g. \citet{Slavin:2017}). While we do not have as yet any observations to confirm the presence of such clumps inside this \ac{SNR}, those observed by ALMA inside the young core-collapse SNR RXJ\,1713.7$-$3946 \citep{SanoRXJ713:2020} at least so far satisfy the above-mentioned conditions.

We also show in Figure~\ref{fig:SEDs} (bottom left and right panels) results after applying modest adjustments to parameters $n$, $B$, $W_p+e$ and $W_e$ to achieve either a dominantly leptonic or hadronic scenario. All of these parameters could be considered feasible for a mature \ac{SNR}. The lower densities $n=5$\,cm$^{-3}$ used in the leptonic-dominant scenario could be realised if one assumes some of the \ac{ISM} gas in the --13 to --1\,km/s velocity range lies just outside of the \ac{SNR} shock. Taking into account that the peak of GeV gamma-ray emission is not spatially coincident with the peak of the gas distribution, the effective density for hadronic interactions could be even smaller. A lower gas density ($n \sim 1$\,cm$^{-3}$ or less) might infer that much of the gas revealed by the HI and CO observations could have been contacted by the \ac{SNR} in recent times, and, it would be more consistent with the low densities used in the evolutionary model (Tab.\,\ref{tab:radpar}, which would define our inter-clump density $n_{\rm icl}$) to estimate the \ac{SNR}'s age. A consequence of this would be a factor 5 to 10 or so increase in $W_p$ and a lowering of the $W_e/W_p$ ratio, but still within a range of acceptable values.

Overall, our spectral energy distribution modelling for the \ac{SNR} suggests contributions from both leptonic and hadronic processes to the GeV emission, and that either a dominantly leptonic or hadronic scenario is feasible while facing some difficulties. A modification to the hadronic component to account for energy-dependent penetration into a clumpy \ac{ISM} might offer a way to match the steeply falling emission below 10\,GeV. The expected cutoff at 5\,TeV energies or thereabouts for electrons in a mature \ac{SNR} (e.g. see Eq.10 of \citet{2020A&A...634A..59B} assuming a shock speed $\sim$300\,km/s and upstream magnetic field of 5 to 10$\mu$G) would however limit the level of $>$1\,TeV IC emission. A model with full-time-dependent energy losses and acceleration would be needed to more accurately predict the electron and proton cutoff energies.   

Future TeV gamma-ray observations by H.E.S.S. and the Cherenkov Telescope Array (CTA) will therefore be crucial in testing hadronic vs. leptonic models, and the maximum energies of the associated particles. It should also be noted that the age of the remnant, which is still poorly constrained, is a crucial parameter in determining the nature of the gamma-ray emission.

\subsection{Gamma-ray emission from a relic Pulsar-wind Nebula?}

We investigated the possibility that the gamma-ray emission from Diprotodon arises from a relic population of electron-positron pairs produced in a \ac{PWN} during the first few millennia of the system and later mixed in with remnant material. Building upon previous work by \citet{2009ApJ...703.2051G}, \citet{Martin:2024} developed a model for the dynamics and radiation from a \ac{PWN} growing inside an \ac{SNR}, including the possibility of particle escape from the \ac{PWN} to the \ac{SNR} and then from the \ac{SNR} to the \ac{ISM}. In this new model framework, a population of relativistic electron-positron pairs builds up and spreads across the whole PWN-SNR system. This is described qualitatively below, and the whole formalism is provided in \citet{Martin:2024}.

In the present case, we consider the possibility that particles escaped from the nebula into the remnant and still trapped downstream of its forward shock at the current age of the system are responsible for the observed gamma-ray emission. The original nebula is thought to have completely dissolved by now, and the pulsar may have escaped the system owing to its natal kick (but see the more quantitative discussion of this scenario at the end of the section), such that this population of trapped electron-positron pairs is all that remains from the \ac{PWN}.

If the \ac{SN} explosion gave birth to a neutron star, a \ac{PWN} rapidly develops at the centre of the \ac{SNR}, fed by relativistic electron-positron pairs and turbulent magnetic field resulting from the conversion of rotational energy of the neutron star. It is bounded on the inner side by the pulsar wind termination shock, where spin-down power conversion occurs (by mechanisms not specified in the model), and on the outer side by a shell of swept-up stellar ejecta, produced as the \ac{PWN} expands (and assumed to be geometrically thin in the model). The dynamics of the \ac{PWN} is controlled by the pressure imbalance between the nebula's interior and the ejecta's pressure immediately ahead of its outer frontier, taking into account the bounding shell inertia.

In the model of \citet{Martin:2024}, particles in the \ac{PWN} experience a combination of advection and resonant scattering with magnetic turbulence (assumed to be Alfvenic), which is described as an isotropic diffusion process in position. Particles can escape the nebula over a typical time scale set by the time needed to cross the radius of the \ac{PWN} at any given age. The process is energy-dependent and statistically favours the escape of the highest-energy pairs first. These escaping particles then enter the \ac{SNR} volume, where they will be confined by the magnetic barrier that develops at the forward shock (which is a prerequisite for diffusive shock acceleration, ensuring that particles remain in the accelerator for long enough). This magnetic barrier can be characterised by the maximum momentum of the particles that can be confined in the remnant, and the latter evolves in time, first increasing linearly with time until Sedov-Taylor transition and then decreasing following a power law \citep{2019MNRAS.490.4317C}. As a result, at ages past the Sedov-Taylor transition, the remnant is progressively depleted of its highest-energy particles.

For aged systems, the electron-positron pairs originally energised by the pulsar are spread over three zones (the \ac{PWN}, the \ac{SNR} and the \ac{ISM}), and each population has a specific spectral distribution. The corresponding electromagnetic radiation is computed, with IC scattering in the ambient radiation field producing gamma-ray photons, while synchrotron radiation is responsible for emission in the radio to X-ray band. The radiation field used in IC scattering is the one defined in Section~\ref{sec:SED}. The magnetic field involved in synchrotron radiation is defined for the nebula from the fraction of spin-down power injected as magnetic energy, while it is a free parameter for both the \ac{SNR} and the \ac{ISM} volumes.

In the case of Diprotodon, the model predicts gamma-ray emission matching the observations for the following main parameters\footnote{The small density value used, at the low end of the range of values discussed in previous sections, is actually somewhat dictated by the domain of applicability of the model, which is currently restricted to systems not too deep into the reverse-shock interaction stage \citep[see the discussion in][]{Martin:2024}. Nevertheless, the magnitude of particle escape from the PWN to the SNR is mainly driven by considerations related to the pulsar and its nebula and most of it occurs before reverse-shock interaction, so the scenario proposed here is weakly dependent on the assumed interstellar density.}: (i) an assumed distance of 1.0\,kpc; (ii) for the \ac{SNR}, a mass of 15\,M$_\odot$, the kinetic energy of $5\times10^{50}$\,erg, and an ambient density of 0.03\,H\,cm$^{-3}$; (iii) for the pulsar, an initial spin-down power of $7 \times 10^{37}$\,erg/s, an initial spin-down time scale of 3000\,yr, and a braking index of 3; (iv) for energy injection in the nebula, a fraction of 90\% of the spin-down power going into pairs, injected with a broken-power-law spectrum with index 1.8 and 2.6 respectively above and below the break energy of 50\,GeV and an exponential cutoff at 500\,TeV, while the remaining 10\% are evenly shared between large-scale magnetic field and Alfvenic turbulence with a largest spatial scale of 2\% of the \ac{PWN} radius. The parameters for the remnant and the pulsar are typical of the Galactic population. The initial properties obtained for the pulsar can be translated into an initial period of 55\,ms and a magnetic field of $4.6\times10^{12}$\,G, which agrees very well with the typical values inferred for these parameters from the known pulsar population \citep{Faucher-Giguere:2006,Watters:2011}. Similarly, the properties of the remnant agree well with those inferred from a large sample of Galactic \acp{SNR} \citep{Leahy:2020}.

With such parameters, the \ac{SNR} grows to about 30\,pc at an age of 25\,kyr, at which time it is still in the Sedov-Taylor stage. The reverse shock hits the nebula at about 20\,kyr, and numerical hydrodynamical simulations show that significant mixing, if not full disruption of the nebula ensues \citep[e.g.][]{2017ApJ...844....1K}. At this time, however, most of the particles have already escaped the \ac{PWN} and are trapped within the \ac{SNR} (about $10^{48}$ erg total particle energy). As time goes by, the magnetic barrier at the forward shock weakens, such that only particles with energies below $\sim150$\,TeV remain in the \ac{SNR} at an age of 100\,kyr (under the assumption that particle confinement at the forward shock reached a maximum energy of 1\,PeV at Sedov-Taylor transition, with a subsequent decay in time following a power law with index 2).

\begin{figure}
    \centering\includegraphics[width=\linewidth,clip]{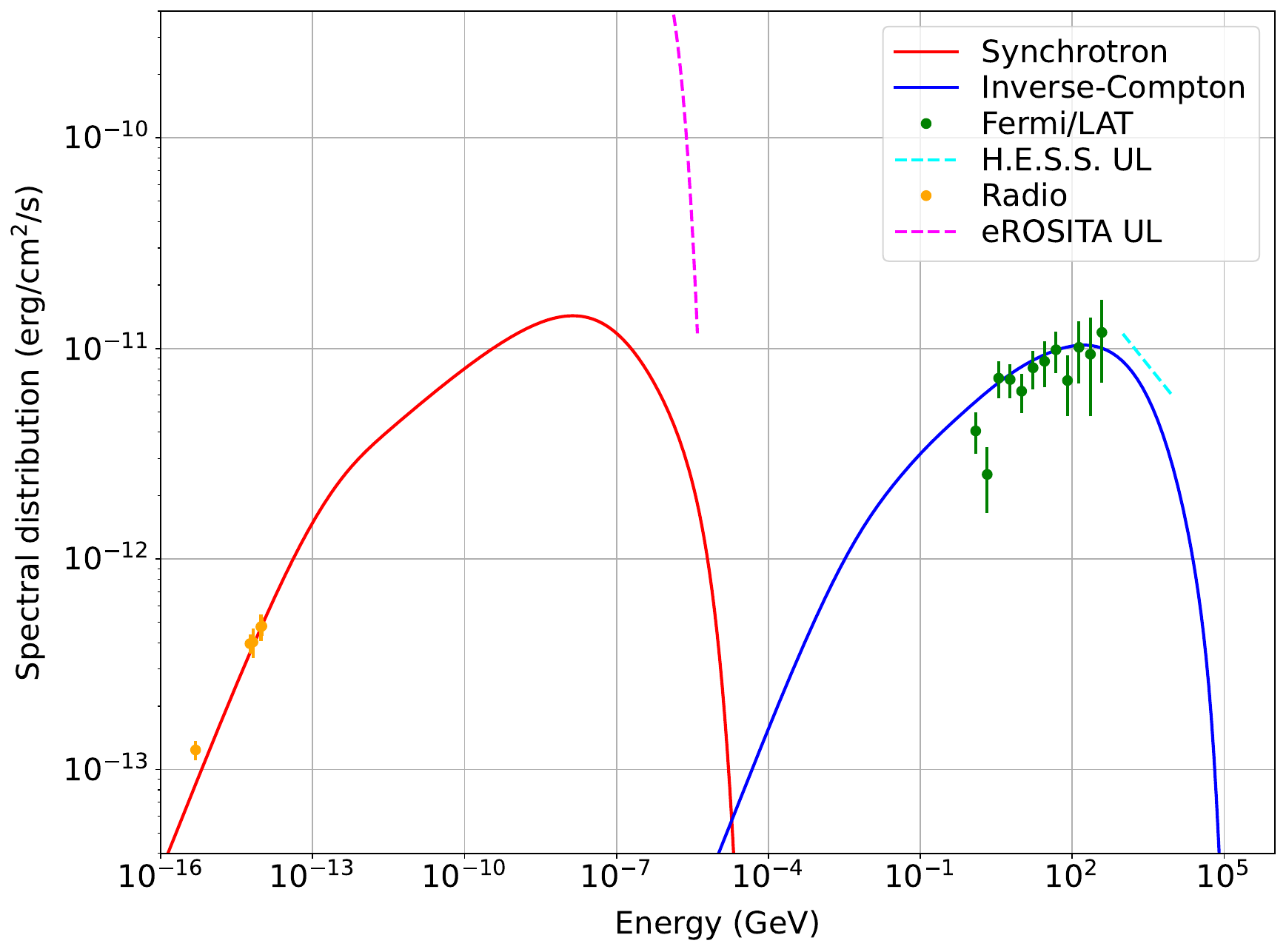}
    \caption{Predicted broadband spectral energy distribution when interpreting the gamma-ray emission as arising from a relic \ac{PWN}, using the model of \citet{Martin:2024}. The dashed lines indicate upper limits on the emission.}
    \label{fig:PWN}
\end{figure}

The predicted gamma-ray emission from the remnant is displayed in Figure~\ref{fig:PWN}. It matches the signal inferred from Fermi-LAT data regarding flux level and spectral shape, and it is consistent with the upper limit from H.E.S.S. The H.E.S.S. upper limit constrains the cutoff energy of the particle injection spectrum and the magnetic field strength in the remnant. Using 500\,TeV for the former, the GHz radio intensity is saturated for an average magnetic field of $5$\,$\mu$G in the remnant. The corresponding radio synchrotron emission falls short of the flux density measured at 0.1\,GHz, despite using parameters for the particle spectrum that are at the end of the ranges inferred for younger \acp{PWN} \citep[a low-energy power-law index of 1.8, below a break energy of 50\,GeV; see][]{Torres:2014}. This suggests that a second population of particles is responsible for most of the radio emission, a possibility that is supported by the fact that the radio and gamma-ray morphologies overlap only partially. Such a population could, for instance, be composed of particles accelerated at the forward shock and subsequently advected downstream of it, which would account for the steeper spectral index. In that case, and if the average magnetic field in the remnant has a strength lower than $5$\,$\mu$G, the relic population of electron-positron pairs have a subdominant contribution to the synchrotron emission and are observable only through their inverse-Compton gamma-ray emission.

Eventually, the relic \ac{PWN} scenario can account for three observables: (i) gamma-ray emission not being positionally coincident with interstellar gas, a priori discarding a hadronic emission mechanism and favouring a leptonic origin; (ii) gamma-ray emission seemingly being pushed away from the galactic plane, which can be ascribed to the reverse shock crushing of the nebula occurring first on the low-latitude side, as the remnant ran into the gas density gradient of the Galactic plane; (iii) a flat gamma-ray spectrum over $\sim$1GeV-1TeV at a relatively advanced age, not characteristic of evolved \acp{SNR} with ages $50-100$\,kyr.

Validating this interpretation would ideally require identifying the pulsar of the system via its magnetospheric or nebular emission. With the assumed model parameters, this pulsar would have a spin-down luminosity of $8\times10^{35}$\,erg/s and a period of 170\,ms at the assumed age of 25\,kyr. Yet, the pulsar may have moved away from the system as a result of its natal kick, although that involves relatively large velocities \citep[about 1200\,\kms\ to travel $\sim$30\,pc in $\sim$25\,kyr, which is at the high end of the natal kick velocity distribution inferred in][]{2017A&A...608A..57V}. Also, such large pulsar velocities would create an observable radio trail as in several other known systems \citep{2014A&A...562A.122P,2019MNRAS.486.2507A,2024PASA...41...32L}. There are five pulsars in the ATNF pulsar catalogue (version 2.0.0) within 2$^\circ$ of the remnant's centre, and all of them have characteristic ages above 100\,kyr, and spin-down powers of the order of $10^{34}$\,erg\,s$^{-1}$ at most (for PSR~J0954$-$5430). This would tend to disfavour the relic \ac{PWN} interpretation, although one cannot exclude that the required pulsar evaded detection because of unfavourable beaming, or that the emission from the (possibly bow-shock) wind nebula is out of reach of the currently available observations. A close inspection of the eROSITA observations of the remnant and its outskirts is warranted to establish if this scenario deserves further consideration.

\section{Conclusion}
 \label{con}

We present new high-resolution and sensitive radio-continuum images of the Diprotodon Galactic \ac{SNR} \g. Our \ac{ASKAP} 943\,MHz image at the resolution of 15\arcsec$\times$15\arcsec\ (\ac{RMS} of 25\,\ujybm) shows an object with an unusually large angular extent of $\sim$3\fdg3 across.
We suggest that Diprotodon is at $\sim$1\,kpc distance and in the radiative phase of evolution. Coupled with a somewhat unexpected hard \textit{Fermi}-LAT gamma-ray spectrum, this extensive ($\sim$D=57\,pc) object challenges our theoretical predictions about the \acp{SNR} evolution.

\begin{acknowledgement}

This scientific work uses data obtained from Inyarrimanha Ilgari Bundara / the Murchison Radio-astronomy Observatory. We acknowledge the Wajarri Yamaji People as the Traditional Owners and native title holders of the Observatory site. \ac{CSIRO}’s \ac{ASKAP} radio telescope is part of the \ac{ATNF}\footnote{\label{foot:ATNF}\url{http://www.atnf.csiro.au}}. 

This work is based (in part) on observations made with the Spitzer Space Telescope, which was operated by the Jet Propulsion Laboratory, California Institute of Technology under a contract with NASA.

We thank an anonymous referee for comments and suggestions that greatly improved our paper.
\end{acknowledgement}

\paragraph{Funding Statements:}
Operation of \ac{ASKAP} is funded by the Australian Government with support from the National Collaborative Research Infrastructure Strategy. \ac{ASKAP} uses the resources of the Pawsey Supercomputing Research Centre. Establishment of \ac{ASKAP}, Inyarrimanha Ilgari Bundara, the \ac{CSIRO} Murchison Radio-astronomy Observatory and the Pawsey Supercomputing Research Centre are initiatives of the Australian Government, with support from the Government of Western Australia and the Science and Industry Endowment Fund.

MDF, GR and SL acknowledge \ac{ARC} funding through grant DP200100784. 
N.H.-W. is the recipient of \ac{ARC} Future Fellowship project number FT190100231.
SD is the recipient of an \ac{ARC} Discovery Early Career Award (DE210101738) funded by the Australian Government.
HS acknowledges funding from JSPS KAKENHI Grant Number 21H01136.
DU and BA acknowledge the financial support provided by the Ministry of Science, Technological Development and Innovation of the Republic of Serbia through the contract 451-03-66/2024-03/200104 and for support through the joint project of the Serbian Academy of Sciences and Arts and Bulgarian Academy of Sciences  ``Optical search for Galactic and extragalactic supernova remnants''. BA additionally acknowledges the funding provided by the Science Fund of the Republic of Serbia through project \#7337 ``Modeling Binary Systems That End in Stellar Mergers and Give Rise to Gravitational Waves'' (MOBY).
MA acknowledges support from Universidad de Costa Rica under grant B8267.
RB acknowledges funding from the Irish Research Council under the Government of Ireland Postdoctoral Fellowship program. 
JM acknowledges support from a Royal Society-Science Foundation Ireland University Research Fellowship (20/RS-URF-R/3712).
CBS acknowledges support from a Royal Society Research Fellows Enhancement Award 2021 (22/RS-EA/3810).
SL and BV were supported by the Ministry of Science, Technological Development and Innovation of the Republic of Serbia (MSTDIRS) through contract no. 451-03-66/2024-03/200002 made with Astronomical Observatory (Belgrade).
PM acknowledges financial support by \ac{ANR} through the GAMALO project under reference ANR-19-CE31-0014.
We thank an anonymous referee for comments and suggestions that greatly improved our paper.

\paragraph{Data Availability:}
The data that support the plots/images within this paper and other findings of this study are available from the corresponding author upon reasonable request. The \ac{ASKAP} data used in this article are available through the \ac{CASDA}\footnote{\url{https://research.csiro.au/casda}}. 
This scientific work uses data obtained from Inyarrimanha Ilgari Bundara / the Murchison Radio-astronomy Observatory.



\appendix
\section{Naming \g\ as Diprotodon}
 \label{sec:app}

\begin{figure}
    \includegraphics[width=\columnwidth]{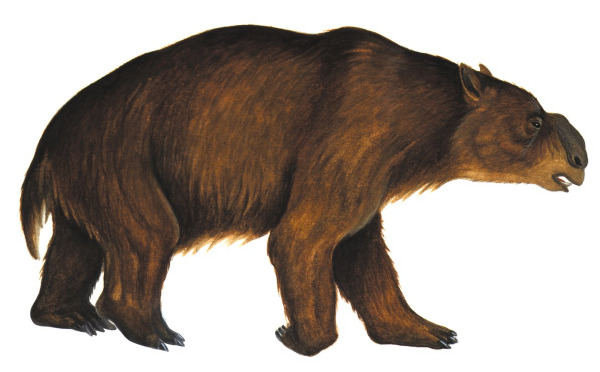}
    \caption{A Diprotodon. Credit: Anne Musser.}
    \label{fig:diprotodon}
\end{figure}

Diprotodon is an extinct genus of Australian marsupial from the Pleistocene period, resembling a giant wombat (see Figure~\ref{fig:diprotodon}). We adopt this name for \g\ to raise awareness of this extinct Australian animal and all of Australia's prehistorical megafauna, as well as to raise awareness of the current extinction rate of several other species in Australia. As \g\ is potentially among the largest \acp{SNR}, it is fitting to liken it to this largest of wombats, Diprotodon.

The sole Diprotodon species, \textit{D. optatum}, was the largest known marsupial, weighing up to 3\,500\,kg and as long as 3.4\,m and 2\,m tall. It became extinct at the end of the continent-wide extinction of megafauna approximately 46\,000~years ago \citep{roberts_2001Sci...292.1888R}. The first fossils were discovered in Adnyamathanha Country in the Flinders Ranges of South Australia, and more recently, archaeologists have found a bone of a juvenile Diprotodon in a dateable Aboriginal deposit inside a rock shelter in the Flinders Ranges. The shelter was unsuitable for access by a live Diprotodon, and the dating was 46\,000--49\,000~years ago \citep{hamm_2016Natur.539..280H}. It is believed that Aboriginal people co-existed with Diprotodon for up to 20\,000 years. Unfortunately, there is no accepted record in Aboriginal rock art clearly showing a Diprotodon \citep{bednarik_2013}. Coincidentally (or not), the Adnyamathanha people have a story about a ``Yamuti'', which was dangerous for children, and if they saw one, they were to climb high on the nearest tree as the Yamuti couldn’t climb \citep{koolmatrie}. Diprotodon, an herbivore, wasn’t a predator; but they also couldn’t climb trees. We now know that Aboriginal stories aren’t just myths but have been shown to be accurate for at least 14\,000~years \citep{Hamacher_2023105819}, so this story could describe living alongside Diprotodon. Perhaps Aboriginal people and Diprotodon alike witnessed the supernova in the night sky which produced \ac{SNR} \g.

\bibliography{DIPROTODON_main}

\end{document}